%% file: main.tex
\documentclass[acmsmall]{acmart}
\AtBeginDocument{%
  }
    
\newcommand{\model}{{\tt MalEval}}
\newcommand{\blackcirc}[1]{%
\tikz[baseline=(char.base)]{
  \node[shape=circle, fill=black, inner sep=1pt] (char) {\textcolor{white}{\footnotesize #1}};
}}
\setcopyright{acmlicensed}
\copyrightyear{2018}
\acmYear{2018}
\acmDOI{XXXXXXX.XXXXXXX}
\acmConference[Conference acronym 'XX]{Make sure to enter the correct
  conference title from your rights confirmation email}{June 03--05,
  2018}{Woodstock, NY}
\acmISBN{978-1-4503-XXXX-X/2018/06}

\usepackage{float}
\usepackage{booktabs}
\usepackage{multirow} 
\usepackage{subfigure}
\usepackage{diagbox}
  
\usepackage{amsmath,amsthm,amssymb,amsfonts}
\usepackage{algorithm}
\usepackage{algorithmic} 
\usepackage{enumitem}
\usepackage{pifont}
\usepackage{bbding}
\usepackage[most]{tcolorbox}
\usepackage[table]{xcolor}
\usepackage{amssymb}
\usepackage{bbm}

\newcommand{\diff}[1]{\textcolor{black}{#1}}
\definecolor{orange}{RGB}{220,92,18}

\setcopyright{cc}
\setcctype{by}
\acmDOI{10.1145/3832187}
\acmYear{2026}
\acmJournal{PACMSE}
\acmVolume{3}
\acmNumber{ISSTA}
\acmArticle{ISSTA096}
\acmMonth{10}
\acmSubmissionID{issta26main-p865-p}
\received{2026-01-30}
\received[accepted]{2026-06-25}

\begin{document}


\title{Is “Knowing It’s Malicious” Enough? Evaluating LLMs for Fine-Grained Malware Behavior Auditing}

\author{Xinran Zheng}
\orcid{0000-0003-1130-7916}
\affiliation{%
  \institution{University College London}
  \city{London}
  \country{United Kingdom}
}
\email{xinran.zheng.23@ucl.ac.uk}

\author{Xingzhi Qian}
\orcid{0009-0005-6617-6521}
\affiliation{%
  \institution{University College London}
  \city{London}
  \country{United Kingdom}
}
\email{xingzhi.qian.23@ucl.ac.uk}

\author{Yiling He}
\orcid{0000-0002-5977-1489}
\authornote{Corresponding Author.}
\affiliation{%
  \institution{University College London}
  \city{London}
  \country{United Kingdom}
}
\email{heyilinge0@gmail.com}

\author{Shuo Yang}
\orcid{0000-0003-1638-9623}
\affiliation{%
  \institution{University of Hong Kong}
  \city{Hong Kong SAR}
  \country{China}
}
\email{shuoyang.ee@gmail.com}

\author{Lorenzo Cavallaro}
\orcid{0000-0002-3878-2680}
\affiliation{%
  \institution{University College London}
  \city{London}
  \country{United Kingdom}
}
\email{l.cavallaro@ucl.ac.uk}





\renewcommand{\shortauthors}{Xinran Zheng et al.}

\begin{abstract}
Automated malware classifiers achieve strong detection performance, but auditing requires more than flagging a sample: analysts must explain malicious behaviors and justify them with concrete code evidence, a requirement that traditional signature-based methods and learning-based XAI often fail to satisfy in a human-interpretable manner. Large Language Models (LLMs) appear well-suited for this task due to their code reasoning and summarization ability, yet it remains unclear whether they can support reliable auditing. In particular, evaluating them faces three hurdles: (1) the lack of detailed, human-written behavior ground truth for reliable benchmarking; (2) real-world application codebases typically exceed the context limits of current models, which cannot be fully processed at once; and (3) the absence of reliable mechanisms to verify whether LLM-generated behavioral claims are faithfully supported by concrete code evidence.

Together, these obstacles make benchmarking LLM-based auditing non-trivial, leaving their true capabilities and failure modes opaque. 
\diff{To bridge this gap, we introduce \model, a diagnostic evaluation framework for systematically measuring the capability boundaries of LLMs in malware auditing.}
We pair real-world application codebases with expert-written audit reports to obtain fine-grained, behavior-level ground truth. Large codebases are compressed into unified behavior-relevant program contexts via a context-driven intermediate representation that preserves essential call relations. Both expert reports and model outputs are then mapped, through constrained reasoning, into structured evidence chains linking code-level facts to high-level behaviors in a common, comparable space. Built on this foundation, \model\ decomposes auditing into 4 stage-wise auditing tasks, allowing each intermediate judgment to be independently verified under limited context windows.
We leverage \model\ to evaluate seven widely used LLMs and uncover clear capability boundaries: models rely on surface cues over verifiable evidence, struggle to compose dispersed facts into coherent attack chains, and are highly sensitive to context formulation. These findings shift the focus from optimizing isolated outputs to designing LLM and agentic workflows that can reliably support malware auditing.

\end{abstract}



\maketitle
\input{Tex/Introduction}

\input{Tex/Review}
\input{Tex/Methodology}

\input{Tex/Evaluation}

\input{Tex/Conclusion}

\input{Tex/DataAvailability}
\bibliographystyle{ACM-Reference-Format}
\bibliography{software}










\end{document}

%% file: Tex/Introduction.tex
\section{Introduction}
\label{sec:intro}
The expansion of the Android ecosystem has intensified malware threats, requiring robust defenses beyond initial detection. While automated systems serve as a critical first line of defense~\cite{Arpdrebin, Grossedeepdrebin, zheng2025learning, hcc, qian2025lamd}, their output merely initiates the operational workflow. Once a sample is flagged as suspicious, it enters the stage of security auditing, where analysts must not only characterize the malware’s behaviors but also substantiate such claims with explicit supporting analyses, which goes beyond behavior summarization: it requires constructing systematic behavior reports grounded in causal evidence, as well as reassessing and refuting prior detections when the evidence does not hold, thereby acting as a corrective safeguard in Security Operations Center (SOC) workflows~\cite{chen2017automated}.

Despite decades of progress, existing malware analysis techniques remain ill-suited for scalable and verifiable auditing, where conclusions must be grounded in concrete program evidence rather than heuristic signals. Signature-based methods remain widely used in industry, but they often yield pattern-level explanations, such as byte-level patterns or cryptographic hashes, that lack the semantic context needed to explain the logic behind malicious actions. Program-analysis-based systems provide richer execution or data-flow evidence, but require substantial expertise and can suffer from limited coverage in large Android applications~\cite{enck2014taintdroid,yan2012droidscope,tam2015copperdroid}. Learning-based interpretation methods~\cite{deepreflect,he2023finer,guo2018lemna} are more scalable, yet they usually operate on coarse-grained model features, provide limited traceability to concrete code evidence, and remain vulnerable to spurious correlations~\cite{kumar2023causal}. Thus, existing paradigms still fall short of scalable, evidence-grounded malware auditing without labor-intensive expert analysis.


Recently, LLMs and agentic frameworks have emerged as a promising direction for malware auditing, owing to their stronger code reasoning and summarization capabilities~\cite{malware_exploring,qian2025lamd,sun2025malloc,anthropic2026mythos}. However, their value for auditing remains insufficiently understood: we still lack a principled way to diagnose where and why LLM reasoning fails along the auditing process, as well as a unified framework for comparing models under controlled conditions. Three challenges make such evaluation non-trivial. \emph{(1) Ground-truth scarcity}: unlike tasks with direct references, such as secure code generation~\cite{he2023large} or malware classification~\cite{he2023msdroid}, malware auditing requires detailed analyst-written reports as ground truth. These reports are costly to produce and rarely available in structured form. \emph{(2) Noise interference}: malicious logic is often buried in large benign codebases~\cite{qian2025lamd}. Feeding raw code easily overwhelms LLMs with irrelevant context~\cite{vstorek2025sense}, while isolated function summaries~\cite{he2025benchmarking} often discard the dependency structure needed for behavioral reasoning. \emph{(3) Evidence verifiability}: high-level malicious claims should be grounded in concrete code evidence for verification, yet the semantic gap between low-level program artifacts and natural-language threat descriptions makes it difficult to distinguish genuine reasoning from plausible guesswork. Existing studies only partially address this gap. Malware-specific efforts either demonstrate feasibility in limited settings~\cite{sun2025malloc} or build multi-stage analysis workflows without fine-grained, expert-aligned evaluation of intermediate reasoning~\cite{qian2025lamd}. General code understanding benchmarks mainly focus on snippet-level functionality or localized programming tasks~\cite{roy2025codesense,xie2025core}, rather than program-level behavioral inference. As a result, we still lack an evaluation framework that aligns code evidence with analyst reports and diagnoses LLM failures across the auditing process.

\diff{Motivated by this evaluation gap, we introduce \model, a reusable evaluation framework for LLM-based malware auditing. Rather than proposing a new auditing system, \model~provides an evidence-verifiable way to assess whether LLMs can perform malware auditing as human analysts do. \model~combines three key components. First, we curate a benchmark of real-world APKs paired with expert-written threat reports, family names and malware categories from well-known security companies as ground truth. 
Second, to mitigate noise interference in large codebases and place models with different context-window limits on a common evaluation interface, we construct a context-driven intermediate representation (CIR) for evaluation, which summarizes functions with local caller-callee context into a bounded input format, allowing different models to be compared under the same code evidence budget.
Third, to bridge the semantic gap between code-level evidence and analyst-written threat reports, we define a constrained output protocol that maps both expert reports and model outputs into a shared evidence$\rightarrow$behavior$\rightarrow$summary$\rightarrow$malicious verdict space with concrete text or code supports. This protocol standardizes observable outputs rather than prescribing how a model should audit malware internally, making intermediate judgments directly comparable and verifiable.}

\diff{Built on this, \model~formalizes malware auditing as four stage-wise evaluation tasks that mirror the analyst workflow identified by Wong \textit{et al.}~\cite{yong2021inside}, based on a user study of 21 analysts from 18 companies: \emph{T1: Function Prioritization}, prioritizing suspicious functions in large codebases; \emph{T2: Evidence Attribution}, attributing low-level indicators to malicious capabilities with explicit evidence; \emph{T3: Behavioral Synthesis}, integrating grounded evidence into behavior-level conclusions; and \emph{T4: Sample Discrimination}, aggregating evidence and behaviors from T1-T3 to reassess samples and distinguish genuine threats from hard-to-separate benign applications. By freezing auditing into a step-wise evaluation lens, \model~makes it possible to diagnose not only whether an LLM succeeds, but also where and why its reasoning breaks down. The resulting protocol is reusable: future studies can keep the same benchmark, task definitions, output units, and metrics while replacing the evaluated model, prompt, context construction strategy, or auditing agent.}

\textbf{Key Findings.}
Using \model\ as a diagnostic lens, we identify three recurring capability gaps in current standalone LLM-based malware auditing. \blackcirc{1} \textit{Evidence recovery and composition remain the primary bottleneck}: Models often recognize locally suspicious code cues, but struggle to recover decisive evidence and assemble dispersed fragments into coherent attack chains; indeed, decisive-evidence missing and attack-chain composition failure together account for over 58\% of auditing errors (Section~\ref{sec:failure_modes}). \blackcirc{2} \textit{Auditing performance is highly sensitive to context construction}: Removing inter-procedural structure degrades evidence attribution by up to 25\% 
and increases hallucination, whereas adding metadata mainly improves final decision-level judgment rather than fine-grained code-grounded reasoning  (Section~\ref{sec:rq2}). \blackcirc{3} \textit{Threat attribution remains the hardest semantic step}: Even when models recover plausible malicious behaviors, they still struggle to determine which behaviors are diagnostically decisive for malware categorization, leading the best category-level F1 reaches only 32.93\% (Section~\ref{sec:rq3}). Together, these findings point to a concrete design direction for next-generation malware auditing systems: behavior-aware context construction, knowledge-guided evidence composition, and taxonomy-grounded threat attribution.

We summarize the contribution of this paper as follows:
\begin{itemize}
    \item \diff{We propose \model, a fine-grained evaluation framework for malware auditing that decomposes the auditing process into explicit intermediate stages, enabling stage-wise and diagnosable evaluation of where and why LLMs fail.}
    
    \item We construct and release a high-quality benchmark of 255 Android applications with 654,188 reachable functions, paired with 62 expert-written malware analysis reports normalized into verified behavior-level ground truth. We also open-source the full evaluation harness to support reproducibility.


    \item We address noise interference and evidence verifiability challenges, the key hurdles in auditing evaluation, via context-driven intermediate representations and constrained structural Chain-of-Thought (CoT) reasoning. This methodology enables the evaluation of LLMs across four analyst-aligned tasks, ranging from function prioritization to behavior synthesis.

    \item Through extensive experiments on 4 open-source and 3 closed-source LLMs, we identify recurring bottlenecks in evidence recovery, attack-chain composition, and threat attribution, yielding practical insights for future LLM-based and agentic malware auditing systems.
\end{itemize}

%% file: Tex/Review.tex
\section{Background}
\subsection{Malware Behavior Analysis to Auditing}

Malware behavior analysis has long been a core problem in Android security, aiming to identify suspicious behaviors from program code or execution traces. Rule- and program-analysis-based methods remain widely used in practice, but they often produce precise yet low-level signals that require substantial expert effort to interpret as malicious intent. They also remain vulnerable to obfuscation, incomplete execution coverage, and conditional triggers, limiting their scalability in operational settings~\cite{arzt2014flowdroid,enck2014taintdroid,yan2012droidscope,tam2015copperdroid}.
To improve scalability, learning-based XAI methods have been proposed for fine-grained malware interpretation. Prior work identifies suspicious snippets~\cite{he2023msdroid}, sensitive functions~\cite{he2023finer}, or isolated behavioral concepts~\cite{he2024dream}. While useful for highlighting suspicious regions, these approaches still produce fragmented indicators rather than coherent reasoning chains from concrete code evidence to high-level malicious behaviors. In addition, these methods either rely on statistical interpretability~\cite{guo2018lemna} or require substantial manual analysis with heuristic refinement~\cite{sun2016poster}, making it difficult to reconstruct a verifiable narrative. More recently, LLM-based analysis has offered a new paradigm by directly interpreting code and generating natural language behavior reports. Early efforts used decision centric prompts with pre-extracted features~\cite{malware_decisioncentric}, while later works explored slicing and layered reasoning for behavioral attribution~\cite{malware_exploring,qian2025lamd,sun2025malloc}. While promising, prior work mainly shows that LLMs can generate plausible behavior summaries, but lacks mechanisms to evaluate whether these claims are grounded in verifiable code evidence. As a result, generated narratives may reflect semantic shortcuts rather than factually supported analysis.


This gap motivates a shift from malware analysis to malware auditing. While analysis identifies potential behaviors, auditing imposes stricter requirements: analysts must ground behavioral claims in verifiable program evidence, synthesize scattered indicators into coherent intent, and reassess upstream detections to filter false positives. Thus, auditing extends behavior analysis with verification, attribution, and corrective judgment. Whether current LLM-based techniques can meet these requirements at scale remains underexplored.

\subsection{Assessing LLMs for Security-centric Code Understanding}
Numerous benchmarks have been proposed to evaluate LLMs on code understanding~\cite{hou2024large,xie2025core,chen2024reasoning,roy2025codesense}. HumanEval and CodeXGLUE~\cite{chen2021Humaneval,lu2021codexglue} measure correctness on isolated snippets, while more recent benchmarks, including CoRe~\cite{xie2025core}, REval~\cite{chen2024reasoning}, CodeSense~\cite{roy2025codesense}, and A.S.E.~\cite{lian2025ase}, move toward richer semantics and repository-level evaluation, sometimes focusing on the security of generated code. However, these evaluations are largely designed for software engineering settings with pre-refined inputs and deterministic correctness.
In contrast, malware auditing poses different requirements: (i) verifiable grounding of claims in concrete program evidence~\cite{he2023finer}; (ii) noise-resilience within large, obfuscated codebases\cite{wang2024llmdfa,nikiema2025code}; and (iii) compositional reasoning to infer high-level intent from fragmented implementations~\cite{wang2022malradar}. Existing benchmarks only partially satisfy these principles, leaving a gap for security-critical reasoning. Preliminary efforts such as CAMA~\cite{he2025benchmarking} represent an encouraging step toward incorporating LLMs into malware behavior analysis through scalable, function-summary-based evaluation. However, while CAMA lets LLMs perform behavior analysis, it fails to ground evaluation on verifiable ground truth and constrained code space, making results expensive and hard to attribute to model ability. \diff{TraceRAG~\cite{zhang2025tracerag} focuses on malware detection rate and analyzes behavior reports on a small manually assessed set, where human reviewers mainly judge the overall quality and usefulness of the final reports. While valuable, this evaluation does not provide fine-grained verification and depends on subjective understandings, hindering fair comparison.}
Consequently, the capability boundaries of LLMs in fine-grained malware auditing remain unclear, highlighting the need for an analyst-aligned evaluation framework to diagnose model limitations and inform the design of future LLM-based agentic auditing systems.

%% file: Tex/Methodology.tex
\section{MalEval}
\label{sec: maleval}
\subsection{Overview}
\begin{figure}[t]
    \centering
    \includegraphics[width=1.0\linewidth]{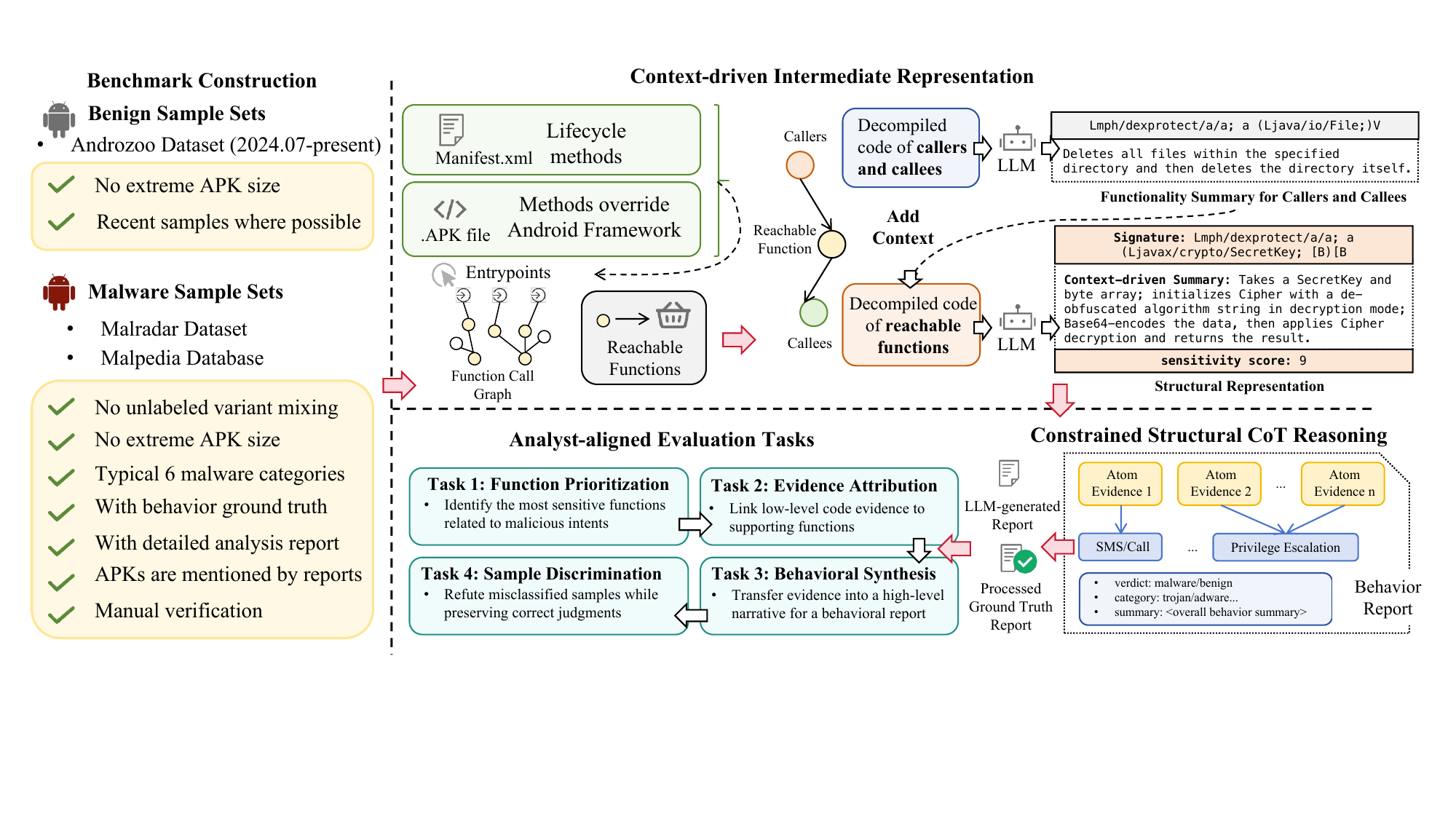}
    \caption{Evaluation pipeline of \model.}
    \label{fig:arch}
\end{figure}

\model\ is a diagnostic evaluation framework for probing LLM capability boundaries in malware behavior auditing. Rather than improving end-to-end audit generation, it makes the auditing process decomposable, comparable, and evidence-verifiable by breaking it into explicitly checkable reasoning stages. The framework comprises three modules, as shown in Figure~\ref{fig:arch}.

\textit{Context-driven Intermediate Representation Construction.}
This module transforms each application's decompiled code and reachable function call graph into a standardized function-level evaluation interface. By distilling caller-callee contexts and potential risks into compact units via evaluated LLMs, CIR supports evaluation on a bounded but behavior-relevant code substrate and reduces noise interference in large codebases.

\textit{Constrained Structural Chain-of-Thought Reasoning.}
This module aligns low-level code understanding with the semantics of expert-written reports by mapping both model outputs and ground-truth reports into a shared \textit{evidence-behavior-verdict} space. The key abstraction is an atomic evidence triplet, \(\langle\)action, asset, target\(\rangle\), which captures a minimal malicious operation in a form that is both code groundable and semantically comparable across reports. Using these triplets as the common interface, this module extracts atomic evidence, aggregates evidence into normalized behaviors, and synthesizes final verdicts, making intermediate reasoning steps directly comparable and verifiable across code and report domains.



\textit{Fine-grained Analyst-aligned Evaluation.} 
Built on the shared representation, this module operationalizes malware auditing as four measurable stages mirroring the core analyst workflow:
\textit{(1) Function Prioritization} tests whether a model can identify security-critical functions for inspection;
\textit{(2) Evidence Attribution} evaluates whether code-grounded atomic evidence is correctly extracted and linked to malicious semantics;
\textit{(3) Behavioral Synthesis} measures whether dispersed evidence can be composed into coherent behavior-level conclusions; and
\textit{(4) Sample Discrimination} assesses whether the resulting audit can reject benign false alarms while preserving true threats.
With stage-specific metrics, \model\ turns malware auditing into a failure-localizing evaluation problem, revealing not only whether a model succeeds, but which reasoning stage breaks down.

\subsection{Benchmark Construction}
\label{sec:benchmark_construction}
\diff{To enable systematic and realistic evaluation of LLMs for malware behavior auditing, we construct a benchmark that pairs Android applications with expert-written malware behavior reports as reference annotations, reflecting practical threat-research workflows where analysts reason over complex application artifacts and synthesize their findings into narrative reports. Table~\ref{tab:category} summarizes the statistics of the curated benchmark.} 

\subsubsection{Application Collection}
\label{sec:app_collection}
\diff{We curate a total of 255 Android applications, comprising 230 malware samples and 25 benign applications, the latter serving as simulated false positives from upstream detectors. After static reachability analysis, these applications yield 654,188 reachable functions (reduced from $\sim$3M total functions), forming the evaluation substrate for auditing. The full evaluation pipeline processes over 2.3 billion input tokens across all models.}

\paragraph{Malware Set}
\label{sec:malware_set}
\diff{Due to the scarcity of high-quality Android malware reports paired with downloadable APKs, we construct our benchmark from two complementary public sources: MalRadar~\cite{wang2022malradar}, which provides Android-specific behavior taxonomies and family annotations, and Malpedia~\cite{malpedia}, which supplements the benchmark with expert-reported and recent samples. Starting from 4,534 MalRadar samples, we exclude samples whose available reports mix unlabeled variants or lack clear variant-level grounding to avoid ambiguous report-to-sample alignment. We then retain six dominant Android malware categories: adware, spyware, trojan, banker, rootkit, and ransomware, and keep only samples with detailed single-family analyst reports whose APKs are downloadable from Koodous\footnote{\url{https://koodous.com}}. This choice ensures reliable family-level grounding in the current benchmark, while future extensions can incorporate available variant-level reports to evaluate behavioral differences within the same family. We further remove redundant APKs only when their complete \texttt{.dex} file sets are identical, considering both single-dex and multi-dex apps. Since our static front end analyzes the Dalvik codebase, identical \texttt{.dex} sets lead to the same extracted functions, call contexts, and function call graphs. APKs sharing only part of their \texttt{.dex} files are retained.}

\diff{Additionally, we exclude oversized or failed apps, and downsample over-represented categories to control evaluation cost and prevent dominant categories from biasing aggregate performance, yielding 200 MalRadar samples. We then add 30 recent expert-reported Android samples from Malpedia where possible, although only a small fraction of its entries satisfy our report-quality and APK-availability requirements. The final malware set contains 230 samples spanning 56 families across six categories. Malware family labels are inherited from the source datasets, while behavior categories follow the MalRadar taxonomy; for the 30 Malpedia samples, we apply the same report-based annotation procedure and validate the labels with four PhD-level researchers.}

\paragraph{Simulated False Positive Benign Set}
\diff{In realistic SOC pipelines, upstream detectors flag entire APKs as suspicious, and analysts audit at the application level to confirm or dismiss alerts. We therefore include complete benign applications as realistic auditing inputs in SOC pipelines to simulate false positives produced by upstream malware detectors. Motivated by concept drift~\cite{transcending}, detectors trained on historical data can misclassify newly released benign applications, forwarding them to the auditing stage as false alarms. To approximate this scenario, we include 25 benign Android applications released after July 2024 from AndroZoo\footnote{\url{https://androzoo.uni.lu}}. The benign proportion of approximately 10\% is consistent with false positive rates reported by state-of-the-art detectors~\cite{Arpdrebin}, enabling evaluation of whether LLMs can correctly dismiss false alerts.} 

\paragraph{Benchmark Statistic}
\diff{We provide descriptive statistics for the curated benchmark in Table~\ref{tab:category}. Malware family labels are inherited from expert reports, while malware category labels follow the MalRadar~\cite{wang2022malradar} taxonomy; for the additional Malpedia~\cite{malpedia} samples, we apply the same report-based annotation procedure and validate the resulting labels with four PhD-level researchers. The table summarizes category and family distributions for the malware and benign sets, and also reports protection characteristics, including packing and obfuscation. We automatically annotate these protection characteristics using Androguard~\cite{androguard} with heuristic rules. A sample is marked as \textit{Packed} if it matches known packer signatures from public rule sets~\cite{apkid_github}, or if static analysis reveals patterns that decrypt or unpack hidden DEX code before loading it via Android dynamic code-loading APIs such as \texttt{DexClassLoader} or \texttt{DexFile}. Obfuscation is assessed only for non-packed samples, since packing already hides the main payload and dominates the analysis difficulty. A non-packed sample is marked as \textit{Obfuscated} when more than 50\% of its identifiers are short names or when it contains high-entropy strings indicative of name mangling or string encoding.}

\input{Tab/category}

\subsubsection{Expert Report Acquisition and Normalization}
\label{para:report_parsing}
\diff{Each malware sample is linked to an expert-written behavior report from major security vendors, using sample hashes to establish reliable alignment between reports and APKs. Because analyst reports are often written at the family level, multiple APK samples from the same family may correspond to a single report; therefore, the number of reports is smaller than the number of malware samples.}

\diff{To remove irrelevant page content, we apply lightweight normalization: raw HTML reports are converted to PDFs and strip advertisements, scripts, and navigation elements, retaining only substantive analysis pages. In total, we process 62 expert reports for all 230 malware samples, which are then transformed into structured ground truth containing behavior labels, supporting evidence, and high-level summaries for evaluation.}

\subsection{Context-driven Intermediate Representation Construction}
\label{sec: intermediate representation}

\subsubsection{Code Context Reduction}

To focus analysis on behavior-relevant code and improve efficiency, we first restrict each application's codebase to a reduced function domain $\mathcal{F}_r$, over which all subsequent representation construction and evaluation are performed. Starting from identified entrypoints, we construct the function call graph and extract all reachable functions via Breadth-First Search (BFS), effectively filtering out dead code and reducing token overhead in auditing.

\diff{We adopt a FlowDroid-style static entrypoint model~\cite{arzt2014flowdroid}, which treats Android lifecycle methods and framework callbacks as execution roots. Entrypoints are identified by (i) extracting lifecycle and callback methods declared in the Android Manifest, and (ii) detecting user-defined methods that override or implement Android framework APIs through class hierarchy analysis. This approach provides broad coverage of executable behaviors under a static setting and is widely used in large-scale Android analysis. We use this static front-end as a scalable and reproducible instantiation of context construction, rather than as a requirement of \model\ itself. The evaluation protocol only requires a set of behavior-relevant functions and their contextual relations; these inputs may also be produced by alternative static analyzers, dynamic traces, hybrid analysis, or retrieval-based agentic workflows. In our implementation, this process yields 801,065 analyzed functions including 654,188 reachable functions across the benchmark, reducing the analysis scope by 78.92\% from the original 3,800,163 functions and making systematic evaluation practical.}

\subsubsection{Context-driven Intermediate Structural Representation}
\label{subsec:context-rep}
Instead of asking models to search over raw application code at scale, which is impractical under limited context windows and may introduce inconsistent context selection, we introduce a context-driven intermediate structural representation, $CIR(f)$, for each reachable function $f \in \mathcal{F}_{r}$. CIR is designed to expose behavior-relevant context within a bounded input budget while grounding model reasoning to concrete functions and their semantic roles. By compressing code into function-level units with local caller-callee context and sensitivity scores, CIR provides a stable, controllable, and reproducible evaluation interface. This design lets different models reason over the same code evidence substrate, rather than relying on model-specific search or truncation over a broad code space.

To construct $CIR(f)$, we derive a context-driven description $\mathcal{D}_{c}(f)$ by incorporating limited structural context from the call graph. Specifically, we consider the one-hop caller-callee neighborhood $\mathcal{C}(f)$ of $f$ to balance function coverage and dependence depth. For each neighboring function $f_{call} \in \mathcal{C}(f)$, we first generate a standalone functionality description $\mathcal{D}(f_{call})$ from its own source code, without contextual information. An LLM with fixed prompts and decoding settings then integrates these neighboring descriptions with the signature $\mathcal{N}(f)$ and decompiled code of $f$, producing $\mathcal{D}_{c}(f)$, a concise description of the role of $f$ within its invocation context. In addition, we associate each function with a sensitivity score $\rho(f)$, which estimates the potential involvement of $f$ in security-relevant behaviors, such as handling sensitive data or invoking privileged operations. Importantly, $\rho(f)$ is not a vulnerability verdict, but a coarse-grained risk prior that provides auxiliary guidance for downstream evaluation. The final representation is defined as $CIR(f)=\langle \mathcal{N}(f), \mathcal{D}_{c}(f), \rho(f)\rangle$. This process yields a context-aware summary that captures the function's role within the application, providing focused and traceable input for reliable evaluation.

\begin{figure}[t]
    \centering
    \includegraphics[width=1.0\linewidth]{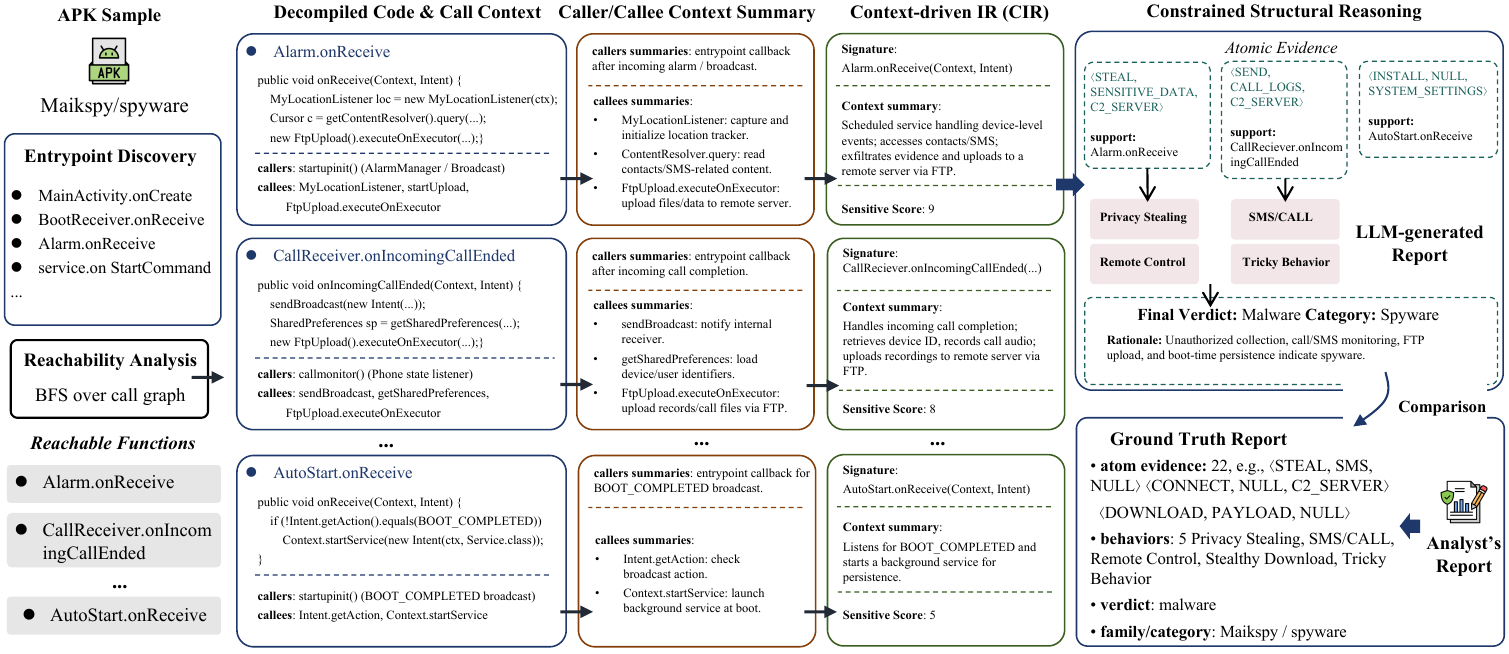}
    \caption{\diff{Running example of \model\ on a Maikspy spyware sample. The figure instantiates Figure~\ref{fig:arch} by tracing one APK from entrypoint discovery to CIR construction, evidence extraction, behavior synthesis, and comparison with expert ground truth.}}
    \label{fig:running_example}
\end{figure}

\input{Tab/atomic}
\subsection{Constrained Structural Chain-of-Thought Reasoning.}
\label{sec:reasoning}
Directly comparing expert reports with LLM outputs is unreliable because free-form natural language cannot be consistently aligned to concrete code artifacts. To make evaluation verifiable, we apply the same constrained, step-wise reasoning to both sides, mapping them into a shared evidence-behavior-verdict space. Evaluation is therefore performed on this common code-grounded representation rather than on unconstrained text.

The core unit of this representation is an atomic evidence triplet, $\langle \textit{action}, \textit{asset}, \textit{target} \rangle$, which captures a minimal malicious operation in a code-grounded and semantically comparable form. To ensure consistency, \model\ defines controlled vocabularies for actions, assets, targets, and behavior labels (Table~\ref{tab:atmoic}). Specifically, we extract verbs and entities from the core analysis text of 42 ground-truth reports, and manually group them into 16 actions, 20 assets, and 13 targets. We further refine these vocabularies on the remaining 20 reports until all observed evidence is covered. The behavior labels follow the MalRadar taxonomy~\cite{wang2022malradar}, where \textit{Miner} is retained as a negative-control label because it belongs to the controlled behavior space but does not appear in the current benchmark.

Based on this schema, reasoning proceeds in three internal stages: (1) extracting atomic evidence from the top-$k$ sensitive CIRs; (2) mapping evidence to normalized behavior categories with explicit support links; and (3) synthesizing a final malware verdict and category, followed by a self-review that checks structural validity, evidence coverage, and schema compliance. 
This design makes the audit traceable by requiring each behavior claim to be supported by atomic evidence and each evidence item to be linked to concrete code units. To convert free-form analyses into comparable structured reports, we apply the same constrained protocol to CIR-based model outputs and expert-written ground-truth reports, using GPT-5~\cite{openai-gpt5-2025} to directly process the original PDF reports for maximal information retention. 

\subsection{Evaluation Pipeline}
\label{sec:example}
\diff{Figure~\ref{fig:running_example} instantiates \model\ on a Maikspy spyware sample and shows how an APK-level audit is transformed into a standardized evaluation instance. Given a flagged APK, \model\ identifies entrypoints, including lifecycle methods and framework callbacks, and performs BFS-based reachability analysis over the Androguard~\cite{androguard} call graph. The reachable functions and their one-hop caller/callee context form the code substrate for evaluation, avoiding direct reasoning over the entire application.
The evaluated LLM participates in the full pipeline rather than only consuming a fixed representation. After function deduplication, it first generates standalone summaries for one-hop callers and callees. These summaries are then combined with each reachable function's signature and decompiled code to construct CIRs as intermediate artifacts produced by the evaluated model itself. For example, \texttt{Alarm.onReceive} is summarized with caller/callee context about contact/SMS access and FTP upload. CIRs are ranked by sensitivity score, and the top-ranked CIRs are sent to the same evaluated LLM to generate a structured audit report through constrained reasoning.
The report contains atomic evidence triplets $\langle action, asset, target\rangle$ with support functions, normalized behavior labels, and a final verdict/category. In the running example, evidence such as $\langle$STEAL, SENSITIVE\_DATA, C2\_SERVER$\rangle$ and $\langle$SEND, CALL\_LOGS, C2\_SERVER$\rangle$ supports behaviors including Privacy Stealing, SMS/CALL, and Remote Control, leading to a spyware verdict. The expert report is normalized with the same structural protocol, enabling comparison at the evidence, behavior, and sample levels in a shared evidence, behavior, and verdict space.}

\subsection{Fine-grained Analyst-aligned Evaluation.}
\label{sec: tasks}
\subsubsection{Evaluation Metrics}
\label{sec:metrics}
To support fine-grained and analyst-aligned evaluation, we define a unified set of metrics spanning three semantic levels: \emph{atomic evidence}, \emph{behavior}, and \emph{sample}. These metrics are task-agnostic and jointly characterize different aspects of auditing quality, from low-level evidence grounding to high-level semantic discrimination.

\begin{figure}[t]
    \centering
    \includegraphics[width=1.0\linewidth]{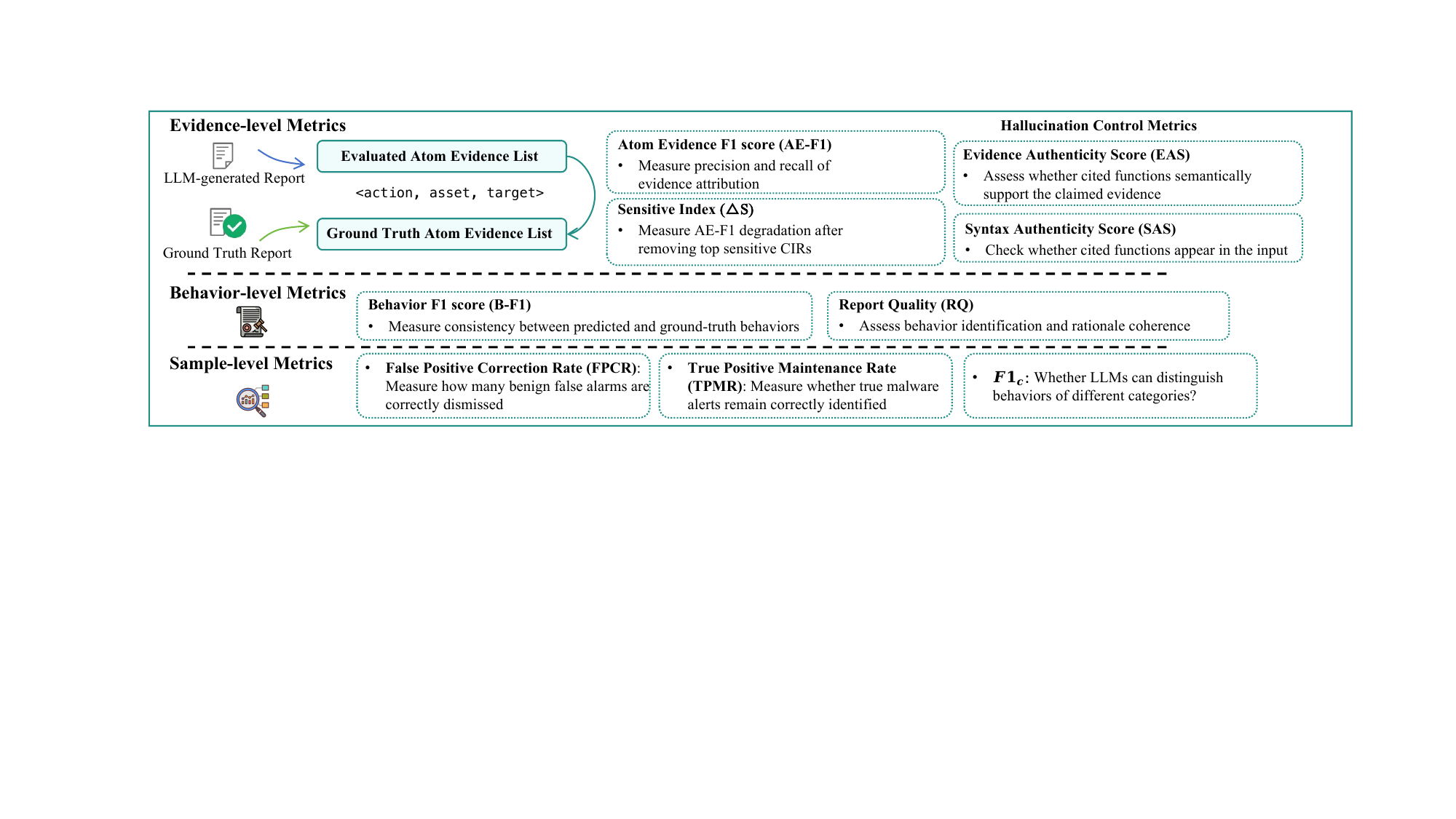}
    \caption{Evaluation Metrics of \model}
    \label{fig:metric}
\end{figure}

\paragraph{Evidence-level Metrics}
Evidence-level metrics evaluate a model's ability to identify and ground the minimal units of malicious evidence. Each atomic evidence item is represented as a structured triplet $\langle \textit{action}, \textit{asset}, \textit{target} \rangle$ (Section~\ref{sec:reasoning}). To tolerate controlled semantic variation, we move beyond exact matching and adopt controlled semantic matching. Component matches are determined by exact equality or predefined equivalence classes over the controlled vocabulary. These classes only merge labels with close security semantics, such as \{\texttt{PREVENT}, \texttt{HIDE}\} and \{\texttt{SMS}, \texttt{CALL\_LOGS}\}. For a predicted evidence item $e_p$ and a ground truth item $e_g$, the alignment score $\mathcal{S}$ is computed as
\begin{equation}
    \mathcal{S}(e_p, e_g) = \omega_1 \cdot s_{action} + \omega_2 \cdot s_{asset} + \omega_3 \cdot s_{target},
\end{equation}
where each component score is 1 for a match and 0 otherwise. We set $\omega_1 = 0.6$, $\omega_2 = 0.3$, and $\omega_3 = 0.1$ to emphasize the role of \textit{action} in behavior interpretation. \textit{AE-F1} is computed with one-to-one matching: each predicted and ground-truth evidence item can be used at most once; pairs with $\mathcal{S}\ge\tau$ are true positives ($\tau=0.8$), and unmatched predictions and ground-truth items are false positives and false negatives. This metric measures evidence-attribution precision and recall while remaining robust to controlled lexical variation. To further quantify reliance on critical code evidence, we introduce the \textit{Sensitivity Index} ($\Delta S$), defined as
\[
\Delta S = AE_{F1}^{full} - AE_{F1}^{retracted},
\]
where $AE_{F1}^{retracted}$ is measured after removing the top 10\% highest-sensitivity CIRs and replacing them with lower-ranked CIRs that were originally excluded, keeping the input size unchanged.

Beyond matching accuracy, we quantify evidence hallucination using two checks: \textit{Syntax Authenticity Score} (SAS), which verifies that all cited functions exist in the input, and \textit{Evidence Authenticity Score} (EAS), which evaluates whether those functions semantically support the claimed evidence. Since function-level malicious labels are unavailable, EAS is computed with an LLM-as-a-judge for code-evidence consistency. Together, AE-F1, EAS, and SAS disentangle evidence identification from grounding authenticity.

\paragraph{Behavior-level Metrics} 
\label{sec:metrics_behavior}
Behavior-level metrics evaluate whether models can correctly synthesize atomic evidence into meaningful malicious behaviors. First, we evaluate behavior consistency using \textit{Behavior F1} ($B_{F1}$), which measures whether the set of inferred behavior labels aligns with the ground-truth behaviors defined under the MalRadar taxonomy~\cite{wang2022malradar}. We also evaluate the quality of the synthesized overall understanding using \textit{Report Quality}(RQ). Conventional metrics like BERTScore~\cite{zhang2019bertscore} often fail in security contexts as they prioritize lexical similarity over causal correctness. Instead, we employ an LLM-as-a-judge paradigm. Each judge evaluates the report against structured ground truth across three dimensions: (i) Objective Identification: captures the core attack goal; (ii) Narrative Coherence: assesses the logical flow of the rationale; and (iii) Logical Derivability: verifies if conclusions stem from identified mechanisms.





\paragraph{Sample-level Metrics} 
\label{sec:metrics_sample}
Sample-level metrics evaluate the reliability of final auditing decisions in realistic operational settings. We measure this from three complementary aspects: (1) \textit{False Positive Correction Rate} (FPCR), which measures the model's capacity to dismiss benign false alarms, reflecting its critical judgment in filtering non-malicious inputs; (2) \textit{True Positive Maintenance Rate} (TPMR), the ability to preserve correct malware judgments without introducing new false negatives; and (3) \textit{Category F1-score}($F1_c$), the macro F1 across malware categories reflecting fine-grained threat understanding. In our evaluation, both benign alarms and confirmed malware verdicts are processed through the \model\ framework, with final verdicts and categories systematically derived from structured reports via constrained CoT reasoning. Together, these metrics disentangle whether a model's performance stems from robust behavioral reasoning or a superficial tendency to label all suspicious inputs as malicious.

\subsubsection{Analyst-aligned Tasks}
To reflect real-world malware auditing, we define four expert-guided tasks, each capturing a distinct stage of the analyst workflow and evaluated with the metrics in Section~\ref{sec:metrics}. \textbf{Task 1: Function Prioritization}
\label{sec: task_1}
In practice, analysts start from a small set of security-sensitive touchpoints rather than inspecting the entire codebase. This task tests whether a model can focus on the small subset of security-critical functions that drive malicious behavior. We measure this using $\Delta S$: a large drop after removing top-sensitive functions indicates that reasoning depends on high-value code evidence rather than prior-driven shortcuts. \textbf{Task 2: Evidence Attribution}\label{sec:task_2} This task evaluates whether a model can identify atomic evidence and ground it to supporting functions. We use \textit{AE-F1} to measure evidence-attribution accuracy, while \textit{EAS} and \textit{SAS} quantify evidence hallucination by checking whether cited support is semantically justified and whether the referenced functions actually appear in the input. \textbf{Task 3: Behavioral Synthesis}\label{sec:task_3} Auditing requires composing dispersed evidence into a coherent attack narrative. This task measures whether a model can synthesize evidence into behavior-level conclusions. We evaluate this using \textit{B-F1} for taxonomy-level consistency and \textit{RQ} for causal coherence quality. \textbf{Task 4: Sample Discrimination}\label{sec:task_4} This task evaluates the model's reliability as a downstream auditing layer that must reject benign false alarms without missing genuine threats. We use \textit{FPCR} to measure false-positive correction, \textit{TPMR} to measure true-positive preservation, and $F1_c$ to evaluate category-level threat attribution.

%% file: Tab/category.tex
\begin{table}\small
\centering
\caption{Information on malware sets, including categories, family counts, obfuscated/packed samples, and average sample size per category.}
\label{tab:category}
\resizebox{0.7\linewidth}{!}{
\begin{tabular}{cccccc} 
\toprule
Category  & Overall & Number of Family  & Obfuscated & Packed & Avg. Size/MB \\ 
\midrule
Trojan     &   27      &    8    &   12              &  7   & 3.82   \\
Banker     &   59      &  20  &  18                   &   7     & 2.77    \\
Rootkit    &   32      &  4     &  0               & 12    & 3.39   \\
Spyware    &  52     &  11   &  13 &  4  & 3.61 \\
Adware     &  25    &  6        &  7            &  11     &  5.55 \\
Ransomware &  35     &  7    &  12              &  0    & 1.71   \\
\midrule
Malware      &   230       &  56   &  62               &   41   & 3.48    \\
\bottomrule
\end{tabular}
}
\end{table}

%% file: Tab/atomic.tex
\begin{table}
\centering
\caption{\diff{Controlled vocabularies for structured evidence and behavior representation, including predefined action, asset, target, and behavior categories.}}
\label{tab:atmoic}
\resizebox{0.7\linewidth}{!}{
\begin{tabular}{l p{1.0\linewidth}}
\toprule
 & \textbf{Element} \\
\midrule
Action & \texttt{STEAL}, \texttt{DOWNLOAD}, \texttt{INSTALL}, \texttt{HIDE}, \texttt{OVERLAY}, \texttt{CLICK}, \texttt{ENCRYPT}, \texttt{CONNECT}, \texttt{PREVENT}, \texttt{REQUEST}, \texttt{GRANT}, \texttt{SEND}, \texttt{INJECT}, \texttt{MONITOR}, \texttt{CAPTURE}, \texttt{EXPLOIT} \\

Asset  & \texttt{CREDENTIALS}, \texttt{FINANCIAL\_DATA}, \texttt{SMS}, \texttt{CALL\_LOGS}, \texttt{MEDIA}, \texttt{LOCATION}, \texttt{DEVICE\_INFO}, \texttt{NOTIFICATIONS}, \texttt{CLIPBOARD}, \texttt{CONTACTS}, \texttt{KEYSTROKES}, \texttt{SENSITIVE\_DATA}, \texttt{APP}, \texttt{PAYLOAD}, \texttt{ROOT\_PRIVILEGES}, \texttt{ADMIN\_PRIVILEGES}, \texttt{CODE}, \texttt{UI\_ELEMENT}, \texttt{COMPUTING\_RESOURCES}, \texttt{NULL} \\

Target & \texttt{FINANCIAL\_APP}, \texttt{SOCIAL\_APP}, \texttt{SYSTEM\_SETTINGS}, \texttt{C2\_SERVER}, \texttt{AD\_NETWORK}, \texttt{USER\_INTERFACE}, \texttt{SECURITY\_SOFTWARE}, \texttt{ACCESSIBILITY\_SERVICE}, \texttt{BROWSER}, \texttt{DEVICE\_ADMIN}, \texttt{HARDWARE\_SENSOR}, \texttt{FILE\_SYSTEM}, \texttt{MINING\_POOL}, \texttt{NULL} \\
Behavior& \texttt{Privacy\_Stealing}, \texttt{SMS/CALL}, \texttt{Remote\_Control}, \texttt{Bank\_Stealing}, \texttt{Ransom}, \texttt{Ads}, \texttt{Miner}, \texttt{Abusing\_Accessibility}, \texttt{Privilege\_Escalation}, \texttt{Stealthy\_Download}, \texttt{Tricky\_Behavior}, \texttt{Premium\_Service}\\
\bottomrule
\end{tabular}
}
\end{table}

%% file: Tex/Evaluation.tex
\section{Experiment}
\label{sec: experiment}
We structure our evaluation around research questions that dissect LLM capabilities in malware auditing. Instead of viewing auditing as a single end task, we break it into targeted probes to reveal what models can do, what evidence they use, and where their reasoning fails.

\begin{itemize}
\item \textbf{RQ1 :}
How effectively do prominent LLMs perform across the four core auditing tasks, and to what extent do they establish a reliable baseline for complex malware auditing scenarios?

\item \textbf{RQ2 :}
How do controlled information perturbations affect behavior reasoning and decision robustness?

\item \textbf{RQ3 :}
Do LLMs exhibit systematic biases or blind spots across different malicious behaviors and malware categories, and where do semantic barriers emerge in behavior understanding?
\end{itemize}

Beyond quantitative metrics, we qualitatively categorize model reasoning failures and use representative cases to show how LLMs fail to bridge the gap between fragmented code facts and high-level malicious intent.

\subsection{Experiment Setting}
\subsubsection{Model Selection}
We evaluate a diverse set of state-of-the-art LLMs, covering both open-source and commercial offerings, as well as general-purpose and code-specialized variants: \textit{Open-source models:} Llama-3.1-8B-Instruct~\cite{meta-llama3.1-2024}, Qwen2.5-Coder-14B-Instruct~\cite{hui2024qwen2}, Qwen3-32B~\cite{yang2025qwen3}, and DeepSeek-R1-0528~\cite{guo2025deepseek}. These represent recent advancements in open LLMs, with Qwen2.5-Coder and DeepSeek specifically optimized for reasoning and code understanding, allowing us to benchmark the progress of open models against commercial APIs. \textit{Commercial models:} GPT-4o-mini~\cite{openai-gpt4o-mini-2024}, Gemini-2.5-Pro~\cite{comanici2025gemini}, and Claude-3.7-Sonnet~\cite{claude37}. These are leading proprietary LLMs known for their strong general-purpose reasoning, robustness, and efficiency. Evaluating them provides a reference point for the upper bound of practical deployment. This selection compares reasoning and non-reasoning models across both open- and closed-source ecosystems, reflecting real-world choices when deploying LLMs in security pipelines.


\subsubsection{Experiment Setting}
\label{sec:exp_setting}
\paragraph{Implementation Details} 
\diff{For open-source models without official APIs, we load the latest HuggingFace checkpoints and serve them locally with vLLM~\cite{kwon2023efficient}: Llama-3.1-8B and Qwen2.5-14B run on one NVIDIA H100 NVL GPU, and Qwen3-32B runs on four NVIDIA A100-40GB GPUs. API-based models, including DeepSeek-R1, GPT-4o-mini, Gemini-2.5-Pro, and Claude-3.7-Sonnet, are accessed through their production endpoints. All evaluations use deterministic decoding with temperature set to 0 and the maximum context window supported by each model. We use the default top-p setting of each endpoint unless the API does not expose it. For each model, CIRs are ranked by sensitivity score and packed in rank order up to the context limit, with a cap of 300 CIRs. This cap standardizes the input budget and enables percentage-based perturbations. We compute $\Delta S$ by removing the top 10\% most sensitive CIRs and recomputing AE-F1. Inference uses up to 16 parallel workers, reduced under API token-per-second or rate limits. All prompt templates, including functionality summarization, CIR construction, structured report generation, ground-truth normalization, and judge scoring, are provided in the artifact.}

\paragraph{Fidelity of LLM-as-a-judge}
\diff{We use an LLM-as-a-judge paradigm to evaluate Evidence Authenticity Score (EAS) and Report Quality (RQ), which require semantic rather than literal match assessment. To reduce single-model bias, we employ three heterogeneous judges excluding the evaluated models, GPT-5~\cite{openai-gpt5-2025}, Gemini-2.5-Flash~\cite{gemini25}, and DeepSeek-V3~\cite{liu2024deepseek}, which independently assess each sample under the same fixed prompt, temperature 0, and deterministic decoding. We quantify inter-judge consistency with average \textit{semantic entropy}~\cite{farquhar2024detecting} over the judges' rationales. Concretely, rationales with embedding cosine similarity exceeding 0.8 are conservatively grouped into one semantic cluster. A lower cluster entropy indicates stronger semantic agreement. Results show that all values in Table~\ref{tab: rq1} remain low (below 0.09).}

\diff{To further validate judge reliability, we manually audited the same 50 randomly selected samples used for the failure mode analysis in Section~\ref{sec:failure_modes}, covering 350 sample-model pairs across seven evaluated models. RQ aligns well with behavior coverage, with only 8 mismatches (2.3\%) between manual inspection and judge scoring. These mismatches arise in borderline cases where a report remains coherent and partially informative despite missing one core behavior.}

\paragraph{Scalability}
\diff{\model\ scales by progressively expanding each APK from entrypoints to reachable functions and then to context-aware semantic summarization. On average, each APK has 747.85 extracted entrypoints, 2,565.44 reachable functions, and 3,303.32 functions after adding one-hop callers/callees and deduplication for functionality summarization. Runtime is dominated by LLM-based semantic processing: static analysis, including entrypoint extraction, reachability analysis, and caller/callee expansion, takes only 0.37 minutes per APK, while DeepSeek-R1~\cite{guo2025deepseek} with 16 workers takes 12.39 minutes for functionality summarization, 17.03 minutes for CIR generation and 1.45 minutes for report generation. The full pipeline takes 31.24 minutes per APK. This wall-clock time may increase under stricter model-serving throughput, token rate, or concurrency limits, but it remains substantially more scalable than manual analysis at a comparable scope.}


\subsubsection{Correctness Checking of Ground Truth.}
To ensure fair comparison, both ground truth and model outputs are processed by the same constrained reasoning framework (Section~\ref{sec:reasoning}). Before evaluation, the structured ground truth is manually audited by four PhD-level security researchers following a unified protocol that checks: (1) the existence of cited raw evidence, (2) correctness of \textit{action}/\textit{asset}/\textit{target} labels, (3) logical support between evidence and behaviors, and (4) consistency between behaviors and the final summary. Across 62 reports, we found and corrected 11 target errors, 8 asset errors, and 9 mislabeled behaviors. All corrections were made by reviewer consensus, and the revised set is used as the final ground truth.


\subsection{RQ1: Effectiveness of LLMs on Malware Behavior Auditing} \label{sec:rq1}
This section answers RQ1 by evaluating seven representative LLMs across four auditing tasks (Section~\ref{sec: tasks}), as summarized in Table~\ref{tab: rq1}. Overall, reasoning-oriented models tend to perform better on malware auditing, but their gains are uneven across stages: stronger models improve behavior-level synthesis more consistently than precise evidence grounding.

\textbf{Obs. 1: Reasoning-oriented models generally achieve stronger auditing performance.}
Models with explicit multi-step reasoning (Qwen3-32B, DeepSeek-R1, Gemini-2.5-Pro, Claude-3.7-Sonnet) consistently outperform weaker or non-reasoning counterparts on evidence extraction and behavior inference. GPT-4o-mini surpasses the code-specialized Qwen2.5-Coder-14B on evidence extraction (AE-F1: 19.12\% vs.\ 12.40\%) despite being a smaller general-purpose model, suggesting that code specialization alone does not determine auditing performance. At the other extreme, Llama-3.1-8B performs poorly across nearly all metrics, with a negligible $\Delta S$ (1.41\%), suggesting its reliance on prior-driven heuristics rather than active code reasoning.

\textbf{Obs. 2: Function-level comprehension does not readily translate into correct evidence composition.} All models achieve substantially higher EAS (58--92\%) than AE-F1 (7-30\%), indicating that they often recognize which functions are semantically related to a behavior, yet still fail to compose these signals into correct evidence triples. For instance, Gemini achieves the highest EAS (92.08\%), but its AE-F1 remains only 24.24\%. This persistent gap suggests that the main bottleneck lies not in isolated function understanding, but in synthesizing cross-function interactions into structured, behavior-grounded evidence, even with relevant context.

\textbf{Obs. 3: No single model dominates all auditing stages, and category attribution remains the weakest stage.} Different models excel at different pipeline stages: Gemini-2.5-Pro achieves the strongest behavior-level synthesis (RQ: 55.67\%, B-F1: 67.04\%), Claude-3.7-Sonnet attains the highest false-positive correction rate (FPCR: 88.00\%), and DeepSeek-R1 provides the strongest balance between evidence extraction and report synthesis (AE-F1: 26.60\%, RQ: 52.57\%). This specialization suggests that auditing stages place different demands on model capability. However, all models share a common weakness in category-level attribution, where the best $F1_c$ reaches only 32.93\%. Thus, even when models recover meaningful behaviors, they still struggle to judge which are most decisive for threat classification. We analyze this behavior-to-category gap in detail in Section~\ref{sec:rq3}.

\input{Tab/RQ1}

\subsection{RQ2: Performance impact of input information perturbations} \label{sec:rq2}
\diff{To evaluate how input context affects auditing, we study two forms of input perturbation. First, we compare CIR with two alternative representations: CIR-decompile, which keeps the same sensitivity ranking and one-hop structure as CIR but replaces descriptions with decompiled code, and LAMD-slice~\cite{qian2025lamd}, which uses backward slices rooted in sensitive APIs (Table~\ref{tab:rep_comparison}). Second, we perturb CIR by (1) removing inter-procedural context and retaining only standalone function descriptions (Table~\ref{tab: rq2}); and (2) adding application-level metadata, including package name, version, declared components, and signing details (Table~\ref{tab: rq3}). Overall, structural code context is necessary for reliable auditing, whereas metadata provides selective gains mainly at the final decision stage.}
\input{Tab/rep_comparison}

\input{Tab/RQ2}
\input{Tab/RQ3}

\diff{\textbf{Obs. 1: CIR balances token efficiency and behavioral coverage as an evaluation interface.} CIR serves as the main input representation for generating behavior reports in \model. To validate its role as an evaluation interface, we compare it with two alternatives: CIR-decompile, which keeps the same function ranking and one-hop structure but replaces CIR descriptions with raw decompiled code, and LAMD-slice~\cite{qian2025lamd}, which uses backward slices rooted in sensitive API callsites. Since LAMD-slice requires sensitive APIs and successful Soot~\cite{soot} analysis, some APKs are excluded; FPCR is therefore marked as ``/'' due to insufficient benign samples. As shown in Table~\ref{tab:rep_comparison}, CIR achieves the highest B-F1 (61.27) and RQ (52.57) with the lowest average token cost and full item coverage. CIR-decompile preserves finer code details and improves FPCR (96.00\%), but its 2.8$\times$ token overhead reduces item coverage to 28.37\%, limiting behavioral coverage and lowering B-F1, RQ and TPMR. LAMD-slice explicitly captures dependencies around sensitive APIs and obtains a slightly higher $F1_c$ (25.94\%), but this API-centered view may miss malicious logic not anchored to sensitive APIs, resulting in the lowest RQ and TPMR. Overall, CIR provides a balanced evaluation interface: it preserves broad behavior-relevant information under a compact token budget while retaining function-level traceability from behavioral claims to code units.}

\input{Tab/category_metric}

\textbf{Obs. 2: Inter-procedural structure is critical for code-grounded auditing.} Removing caller-callee context causes consistent degradation across nearly all tasks. Evidence extraction and behavior identification both drop substantially (AE-F1: $-$1.4 to $-$7.5; B-F1: $-$1.2 to $-$6.6), showing that isolated function summaries are  insufficient for recovering behavior-grounded evidence. The effect is especially strong on function prioritization for the strongest models (e.g., Gemini-2.5-Pro: $-$39.8\%, Claude-3.7-Sonnet: $-$35.7\% in $\Delta S$), indicating that structural links are essential for distinguishing security-critical functions from surrounding noise. FPCR also consistently declines, suggesting that without inter-procedural context, models rely more on isolated suspicious signals. In addition, hallucination increases, as reflected by lower SAS, indicating that models compensate for missing structure by fabricating plausible but unsupported function references.

\textbf{Obs. 3: Metadata mainly improves decision-level judgment rather than fine-grained reasoning.} Adding metadata yields the clearest gains at the sample level, especially for stronger models (e.g., Gemini-2.5-Pro FPCR: 47.8\%$\to$83.3\%, $F1_c$: 28.9\%$\to$33.7\%; Claude-3.7-Sonnet FPCR: 88.0\%$\to$100\%). However, its impact on evidence-level reasoning is mixed: AE-F1 changes are small or inconsistent across models, while $\Delta S$ drops sharply. This indicates that metadata mainly acts as a global decision cue, which may overshadow fine-grained function understanding rather than support behavior-grounded evidence attribution. Its benefits are therefore concentrated at the final judgment stage, while precise evidence attribution remains dependent on structural code context.

\begin{figure}[t]
    \centering
    \includegraphics[width=0.9\linewidth]{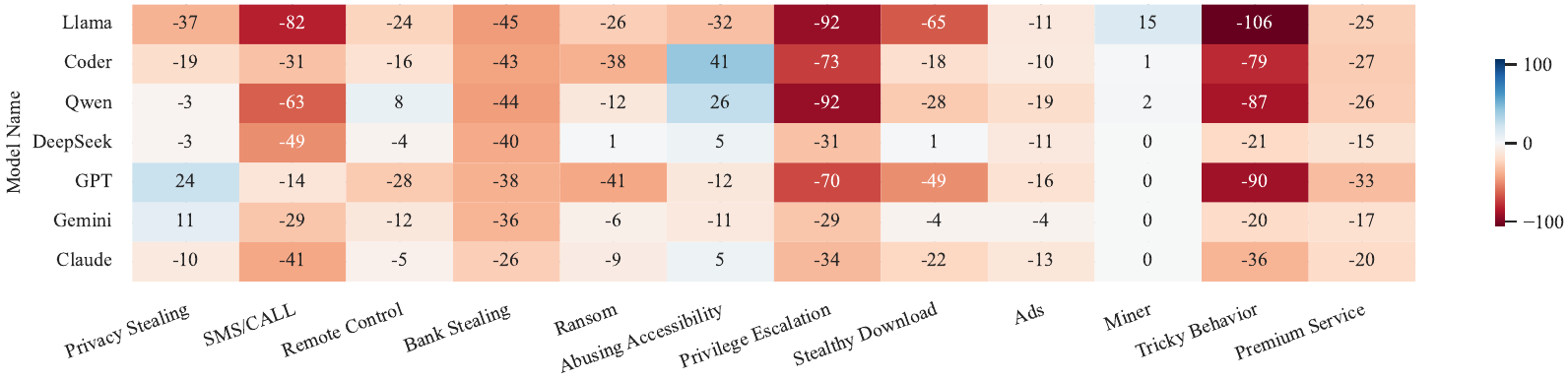}
    \caption{Behavior deviation heatmaps for each LLM trained on malware datasets. Here, shades closer to red indicate the model under-observes that behavior, while blue indicates the model over-observes it. A well-performing model should be closer to white (zero deviation) across all behaviors.}
    \label{fig:heatmap}
\end{figure}

\subsection{RQ3: Performance across Different Behaviors and Categories}
\label{sec:rq3}
We examine how behavior-level prediction tendencies affect malware category attribution. 
Fig.~\ref{fig:heatmap} visualizes deviations between predicted and ground truth behaviors 
(red: under-observation; blue: over-observation), while Table~\ref{tab:misclassification} summarizes category-level failure rates, behavior accuracy (B-F1), report quality (RQ), and detection rates of dominant behavior (B\textsubscript{dom}) for the three strongest models. We define the dominant behavior of each category as the most frequent ground-truth behavior label: \textit{Ads} for \textit{Adware}, \textit{Bank Stealing} for \textit{Banker}, \textit{Ransom} for \textit{Ransomware}, \textit{Privilege Escalation} for \textit{Rootkit}, and \textit{Privacy Stealing} for \textit{Spyware}. We exclude \textit{Trojan} because it spans diverse behavior patterns and lacks a single dominant behavior.

\textbf{Obs. 1: Behaviors with explicit static evidence are recognized more reliably.}
Behaviors tied to explicit Android APIs are recognized more consistently than those requiring compositional or runtime-dependent inference. In Fig.~\ref{fig:heatmap}, behaviors such as 
\textit{Privacy Stealing}, \textit{Abusing Accessibility}, and \textit{Ads} show small deviations across models, as their evidence is often exposed through recognizable APIs or SDK patterns. By contrast, \textit{Tricky Behavior}, \textit{Premium Service}, and \textit{Stealthy Download} are systematically ignored, since they lack stable API signatures and often depend on implicit logic, encrypted strings, or dynamically constructed execution paths. This suggests that behavior recognition is harder when malicious semantics must be inferred from dispersed or runtime-dependent signals, a challenge that is naturally more pronounced under a static evaluation setting.

\textbf{Obs. 2: Category errors stem from weak understanding of category behaviors.}
Table~\ref{tab:misclassification} shows that category prediction depends not only on recognizing malicious behaviors, but on understanding which behaviors are core to each category and how they interact with supporting behaviors. For \textit{Banker} and \textit{Rootkit}, moderate B-F1 but low B\textsubscript{dom} suggests that models recover generic malicious behaviors while missing the core pattern. For \textit{Spyware} and \textit{Ransomware}, the opposite pattern indicates that even when the dominant behavior is detected, models still fail to place it in a complete category context. For example, in a \textit{Rootkit} sample from the \textit{ZNIU} family, DeepSeek identified \textit{Privilege Escalation} but still classified the sample as \textit{Spyware}. This indicates that current LLMs can recognize isolated malicious behaviors, but still lack the category-specific understanding to judge which behaviors are diagnostically decisive for malware categorization. This gap is further reflected in the mismatch between RQ and category correctness: a report may remain coherent and partially informative, yet still miss the behavior most decisive for attribution.

\input{Tab/case_taxonomy}

\subsection{Failure Mode Analysis}
\label{sec:failure_modes}
\diff{To understand how LLM reasoning fails beyond aggregate metrics, we manually analyze 350 generated reports (50 randomly sampled malware applications $\times$ 7 models) and classify each by its primary failure mode. Table~\ref{tab:failure_modes} defines five failure modes ordered along the auditing pipeline, from evidence extraction to threat attribution, with a final mode capturing over-detection. Each incorrect report is assigned to the earliest stage where reasoning breaks down. Concretely, FM-4 (35.63\%) and FM-1 (34.48\%) dominate, followed by FM-3 (23.56\%); together they account for over 93\% of failures. This suggests that the main bottleneck is not local evidence interpretation, but the lack of structured threat reasoning: current LLMs struggle to recover decisive evidence from noisy code, organize it into coherent attack logic, and assign it the correct category-level significance.}

\vspace{3pt}

\diff{\textbf{FM-1: Decisive Evidence Missing.} In this mode, the model fails to recover the decisive evidence needed to identify the sample's core malicious behavior or threat type. As a result, the final judgment relies on generic but weakly discriminative cues, causing the sample to collapse into a broader threat label. In the following example, the final reasoning is dominated by generic \textit{Trojan} cues such as encryption and payload handling, while banker-specific evidence, especially C2-delivered phishing resources and financial-data theft, is absent. Because the decisive evidence for the \textit{Banker} category is not identified early and fails to enter the final reasoning, this case is classified as FM-1.}

\vspace{3pt}
\noindent\begin{minipage}{\linewidth}
\begin{tcolorbox}[
    title=\textbf{FM-1: Decisive Evidence Missing.},
    colback=gray!10,
    boxrule=0pt,
    borderline north={0.5pt}{0pt}{dashed},
    borderline south={0.5pt}{0pt}{dashed},
    left=2pt,
    right=2pt,
    top=2pt,
    bottom=2pt,
    fonttitle=\scriptsize,
    fontupper=\scriptsize,
    fontlower=\scriptsize,
]
\textbf{Generated:}  
Behaviors: \textit{Tricky Behavior}, \textit{Stealthy Download}.  
Key evidence: \{ENCRYPT, SENSITIVE\_DATA, NULL\}, \{DOWNLOAD, PAYLOAD, FILE\_SYSTEM\}.  
\textbf{Category}: trojan.

\tcblower

\textbf{Ground Truth:}  
Behaviors: \textit{Bank Stealing}, \textit{SMS/CALL}, \textit{Privacy Stealing}, \textit{Remote Control}, \textit{Ransom}.  
Key evidence: \{DOWNLOAD, PAYLOAD, C2\_SERVER\}, \{STEAL, FINANCIAL\_DATA, FINANCIAL\_APP\}.  
\textbf{Category}: banker.
\end{tcolorbox}
\end{minipage}

\vspace{3pt}

\diff{\textbf{FM-2: Evidence Misinterpretation.} Here, the model observes code artifacts but reverses their semantic interpretation, typically due to over-reliance on benign-looking cues such as legitimate SDK usage. In the example case, the ground truth contains direct phishing and signals on credential exfiltration, including a fake Instagram login interface and transmission of stolen credentials to a remote server. Instead of treating these as malicious evidence, the model overrides them with benign-looking cues such as \texttt{WebView} and \texttt{AdMob}. This case therefore, reflects misinterpretation of evidence rather than the absence of decisive evidence.}

\diff{\textbf{FM-3: Attack-Chain Composition Failure.} In this mode, the model recovers multiple relevant evidence fragments but fails to connect them into a coherent attack chain. As a result, the final report captures isolated malicious actions yet misses the full threat objective. The model recovers several important fragments in the example, including credential theft and notification monitoring, but fails to organize them into a financial attack chain reflected in the ground truth: fake exchange login, credential capture, OTP theft through notifications, and stealth support for fraudulent transactions. Because the key evidence is present but not integrated into a coherent threat objective, this case is classified as an attack-chain composition failure rather than decisive evidence missing.}

\vspace{3pt}
\noindent\begin{minipage}{\linewidth}
\begin{tcolorbox}[
    title=\textbf{FM-2: Evidence Misinterpretation.},
    colback=gray!10,
    boxrule=0pt,
    borderline north={0.5pt}{0pt}{dashed},
    borderline south={0.5pt}{0pt}{dashed},
    left=2pt,
    right=2pt,
    top=2pt,
    bottom=2pt,
    fonttitle=\scriptsize,
    fontupper=\scriptsize,
    fontlower=\scriptsize,
]
\textbf{Generated Report}:  
\textbf{Behaviors}: (none).  
\textbf{Summary}: The application is a WebView-based app that uses legitimate SDKs for push notifications and analytics. While it collects device information and location data, these behaviors are consistent with standard usage of these SDKs for analytics and targeted notifications. 
\textbf{Verdict}: \textit{benign}.

\tcblower

\textbf{Ground Truth}:  
Behaviors: \textit{Privacy Stealing}, \textit{Tricky Behavior}.  
Key evidence: \{OVERLAY, UI\_ELEMENT, USER\_INTERFACE\}, 
\{STEAL, CREDENTIALS, SOCIAL\_APP\}, 
\{SEND, CREDENTIALS, C2\_SERVER\}.  
Summary: This \texttt{Android/Spy.Inazigram} family masquerades as Instagram boosters, phishes credentials via a fake login UI, and transmits them to remote attacker servers.
\textbf{Category}: spyware.
\end{tcolorbox}
\end{minipage}

\vspace{3pt}

\noindent\begin{minipage}{\linewidth}
\begin{tcolorbox}[
    title=\textbf{FM-3: Attack-Chain Composition Failure.},
    colback=gray!10,
    boxrule=0pt,
    borderline north={0.5pt}{0pt}{dashed},
    borderline south={0.5pt}{0pt}{dashed},
    left=2pt,
    right=2pt,
    top=2pt,
    bottom=2pt,
    fonttitle=\scriptsize,
    fontupper=\scriptsize,
    fontlower=\scriptsize,
]
\textbf{Generated:}  
Behaviors: \textit{Privacy Stealing}, \textit{Tricky Behavior}.  
Key evidence: \{STEAL, CREDENTIALS, NULL\}, \{MONITOR, NOTIFICATIONS, NULL\}.  
Category: \textit{spyware}.

\tcblower

\textbf{Ground Truth:}  
Behaviors: \textit{Privacy Stealing}, \textit{SMS/CALL}, \textit{Bank Stealing}, \textit{Tricky Behavior}.  
Key evidence: \{STEAL, CREDENTIALS, FINANCIAL\_APP\}, \{OVERLAY, UI\_ELEMENT, FINANCIAL\_APP\}, \{SEND, CREDENTIALS, C2\_SERVER\}.  
Category: \textit{trojan}.
\end{tcolorbox}
\end{minipage}

\vspace{3pt}

\diff{\textbf{FM-4: Threat Attribution Failure.}
Here, the model recovers substantial evidence of malicious behavior but fails to assign the correct category-level significance. As a result, the report remains plausible yet maps the sample to the wrong threat category. In the following example, the model recovers a substantial malicious profile, including covert payload delivery and remote communication, but underweights the exploitation capability that defines the \textit{Rootkit} class. The sample is therefore collapsed into the broader \textit{Trojan} category. This case reflects a threat attribution failure rather than a failure in evidence recovery or chain composition.}

\vspace{3pt}
\noindent\begin{minipage}{\linewidth}
\begin{tcolorbox}[
    title=\textbf{FM-4: Threat Attribution Failure.},
    colback=gray!10,
    boxrule=0pt,
    borderline north={0.5pt}{0pt}{dashed},
    borderline south={0.5pt}{0pt}{dashed},
    left=2pt,
    right=2pt,
    top=2pt,
    bottom=2pt,
    fonttitle=\scriptsize,
    fontupper=\scriptsize,
    fontlower=\scriptsize,
]
\textbf{Generated:}  
Behaviors: \textit{Stealthy Download}, \textit{Remote Control}, \textit{Privacy Stealing}, \textit{Tricky Behavior}.  
Key evidence: \{INSTALL, APP, NULL\}, \{DOWNLOAD, PAYLOAD, NULL\}, \{SEND, DEVICE\_INFO, C2\_SERVER\}.  
Category: \textit{trojan}.

\tcblower

\textbf{Ground Truth:}  
Behaviors: \textit{Stealthy Download}, \textit{Remote Control}, \textit{Privilege Escalation}.  
Key evidence: \{DOWNLOAD, PAYLOAD, C2\_SERVER\}, \{INSTALL, PAYLOAD, NULL\}, \{EXPLOIT, NULL, NULL\}.  
Category: \textit{rootkit}.
\end{tcolorbox}
\end{minipage}
\vspace{3pt}

\vspace{3pt}
\noindent\begin{minipage}{\linewidth}
\begin{tcolorbox}[
    title=\textbf{FM-5: Unsupported Extrapolation.},
    colback=gray!10,
    boxrule=0pt,
    borderline north={0.5pt}{0pt}{dashed},
    borderline south={0.5pt}{0pt}{dashed},
    left=2pt,
    right=2pt,
    top=2pt,
    bottom=2pt,
    fonttitle=\scriptsize,
    fontupper=\scriptsize,
    fontlower=\scriptsize,
]
\textbf{Generated:}  
Behaviors: \textit{Remote Control}, \textit{Privacy Stealing}, \textit{Tricky Behavior}, \textit{Stealthy Download}.  
Key evidence: \{CONNECT, NULL, C2\_SERVER\}, \{STEAL, DEVICE\_INFO, NULL\}, \{DOWNLOAD, PAYLOAD, NULL\}.  
Category: \textit{trojan}.
\end{tcolorbox}
\end{minipage}

\vspace{3pt}

\diff{\textbf{FM-5: Unsupported Extrapolation.} In this mode, the model infers malicious capabilities from weak or non-diagnostic signals that do not substantiate such conclusions. Rather than fabricating program elements, the model over-interprets benign Android framework usage as malware evidence, which can lead to false positives. For example, in a benign sample, the generated report maps standard components to unsupported malicious behaviors: \texttt{WorkManager} is interpreted as a \texttt{C2} mechanism, foreground services as evidence of hidden execution, and routine device-state checks as privacy-theft evidence.}

\subsection{Discussion and Threats to Validity}
\label{sec:discussion}
This section discusses the main limitations and threats to validity of the current \model\ instantiation, and clarifies which components are fixed by the evaluation protocol versus replaceable in future extensions.

\diff{\textbf{Analysis Scope and Extensibility}
The current benchmark is built from Android Dalvik bytecode and manifests under a static Android front end. Native code, WebView JavaScript, dynamically loaded payloads, and runtime-only behaviors are outside this front end. These artifacts are not excluded by the protocol itself; once recovered through native-code analysis, dynamic tracing, hybrid analysis, or agentic retrieval, they can be organized into bounded contexts and evaluated with the same protocol.}

\diff{\textbf{Call Graph Dependence}
\model\ does not require a specific call graph algorithm. We instantiate it with a CHA-based static call graph and BFS reachability to obtain behavior-relevant code context. CHA conservatively approximates object-oriented dispatch from declared types and class hierarchy information~\cite{dean1995optimization}, providing broad, deterministic coverage. However, it cannot fully model Android runtime behavior, such as reflection, dynamic loading, native callbacks, and framework dispatch. Different call graphs or dynamic traces may produce different function pools and CIRs. Results should therefore be interpreted under this static context construction setting.}

\diff{\textbf{Reproducibility}
\model\ separates the fixed evaluation protocol from replaceable implementation choices. The benchmark, task definitions, controlled output schema, ground truth reports, and metrics remain fixed, so different models can be compared in the same evidence-behavior-verdict space. In contrast, the evaluated LLM, prompts, and context extraction strategy can be replaced, as long as the resulting inputs follow the CIR format and the outputs follow the controlled schema. This design allows \model\ to be reproduced on the released benchmark and reused with alternative analysis pipelines while keeping the evaluation criteria unchanged.}

\diff{\textbf{Generalizability Beyond Android}
We instantiate \model\ on Android malware as a controlled setting because this ecosystem provides established behavior taxonomies, analyst reports, and downloadable APKs that can be aligned into verifiable behavior-level ground truth~\cite{wang2022malradar}. We acknowledge that Android malware does not represent all security auditing scenarios, especially vulnerability audits of benign desktop, cloud, or server software. Extending \model\ to such settings would change the front end, program representation, taxonomy, and expert reference source. However, the evaluation protocol remains reusable: recovered analysis units can still be represented as CIR-style contexts, normalized into structured evidence and higher-level claims, and evaluated through the same stage-wise tasks and metrics.}

\diff{\textbf{Contamination Considerations}
Contamination is an important concern for LLM evaluation, but rigorous exclusion is difficult here because recent Android samples with both downloadable APKs and expert reports are scarce. Still, \model\ reduces the value of memorization by requiring function-level, CIR-grounded reasoning and intermediate evidence attribution, rather than free-form report matching alone. We therefore treat contamination as a caveat, but not one that accounts for the observed failure patterns.}



%% file: Tab/RQ1.tex
\begin{table}\small
\centering
\caption{Each task and overall performance of LLMs on \model. The input is context-driven intermediate representations (CIR). All results are presented as percentages, and the semantic entropy-based uncertainty scores are shown as subscripts of two LLM-as-a-judge metrics: EAS and RQ.}
\label{tab: rq1}
\setlength{\tabcolsep}{3pt}
\resizebox{0.7 \linewidth}{!}{
\begin{tabular}{c|c|ccc|cc|ccc} 
\toprule
\multirow{2}{*}{Models} & Task 1        & \multicolumn{3}{c|}{Task2}         & \multicolumn{2}{c|}{Task3}                 & \multicolumn{3}{c}{Task4}                                       \\ 
\cmidrule(lr){2-10}
& $\Delta S$ & AE-F1  & EAS  & SAS  & B-F1 & RQ &  FPCR  & TPMR  & $F1_c$   \\ 
\hline
Llama 3.1-8B-I          & 1.41 & 6.94  & $58.28_{\pm 0.08}$  &   69.12  & 36.41    &  $16.74_{\pm 0.05}$    & 40.91    & 78.17    & 11.64    \\
Qwen 2.5-coder-14B-I    &  6.87   & 12.40  & $77.52_{\pm 0.06}$   &  \cellcolor{gray!30}{92.55}      &  49.90  & $31.83_{\pm 0.05}$     & 59.09      & 91.59     & 14.83   \\
Qwen 3-32B              &  8.57    & 19.82    & $82.65_{\pm 0.07}$  &   86.63      &  54.65  & $38.29_{\pm 0.05}$        & 45.83    & 97.76       & 24.62   \\
Deepseek-R1             &  10.45   & 26.60        & $87.49_{\pm 0.07}$  &     90.48     &  61.27  & $52.57_{\pm 0.09}$     & 60.00    & 96.04     & 25.25   \\ 
\midrule
GPT-4o-mini             &  5.89  & 19.12 & $69.42_{\pm 0.08}$  &   77.73          &   55.04 & $32.09_{\pm 0.03}$          & 36.00          & 96.57      & 7.63  \\
Gemini-2.5-pro        &  11.21  &  24.24  & \cellcolor{gray!30}{$92.08_{\pm 0.06}$}  &  90.85  & \cellcolor{gray!30}{67.04}  &   \cellcolor{gray!30}{$55.67_{\pm 0.08}$} &     47.82  & \cellcolor{gray!30}{98.28} & 28.92  \\
Claude-3.7-sonnet       & \cellcolor{gray!30}{12.34}  & \cellcolor{gray!30}{29.95}    & $86.57_{\pm 0.07}$  &   89.28 &     60.46    & $48.68_{\pm 0.09}$             & \cellcolor{gray!30}{88.00}      & 89.69   & \cellcolor{gray!30}{32.93} \\
\bottomrule
\end{tabular}
}
\end{table}

%% file: Tab/rep_comparison.tex
\begin{table}\small
\centering
\caption{\diff{Comparison of input representations. Avg. Tokens reports mean input tokens, and Item Coverage measures the fraction of representation-specific input items retained in the final model input. CIR-decompile keeps the same ranking and one-hop structure as CIR but uses decompiled code, while LAMD-slice~\cite{qian2025lamd} uses backward slices from sensitive APIs.}}
\label{tab:rep_comparison}
\resizebox{0.7 \linewidth}{!}{
\begin{tabular}{cccccccc} 
\hline
Representation & B-F1 & RQ & FPCR & TPMR & $F1_c$ & Avg. Tokens & Item Coverage  \\ 
\hline
CIR            &    61.27  &  52.57  &   60.00   &   96.04   &    25.25    &     18696.85       &         100.0\%     \\
CIR-decompile        &   54.43   &  49.86  &  96.00    & 89.52     &    24.39    &     52158.10        &      28.37\%       \\
LAMD-slice~\cite{qian2025lamd}     &   37.75   &  37.07  &   /   &   71.36   &    25.94    &      50649.80        &      72.94\%      \\
\hline
\end{tabular}
}
\end{table}

%% file: Tab/RQ2.tex
\begin{table}\small
\centering
\caption{Each task and overall performance of LLMs on \model~after removing the context of call-chain from CIRs. Blue and red indicators highlight differences compared to results derived from Table~\ref{tab: rq1}. All results are presented as percentages.}
\label{tab: rq2}
\resizebox{\linewidth}{!}{
\begin{tabular}{c|c|ccc|cc|ccc} 
\toprule
\multirow{2}{*}{Models} & Task 1        & \multicolumn{3}{c|}{Task2}         & \multicolumn{2}{c|}{Task3}                 & \multicolumn{3}{c}{Task4}                                       \\ 
\cmidrule(lr){2-10}
& $\Delta S$ & AE-F1  & EAS  & SAS  & B-F1 & RQ &  FPCR  & TPMR  & $F1_c$   \\ 
\hline
Llama 3.1-8B-I       &  $0.83_{\text{\tiny \color{red} $\downarrow$ 38.41\%}}$  &    $5.31_{\text{\tiny \color{red} $\downarrow$ 23.49\%}}$   &  $52.05_{\text{\tiny \color{red} $\downarrow$ 10.69\%}}$  &       $52.95_{\text{\tiny \color{red} $\downarrow$ 23.39\%}}$    &  $31.65_{\text{\tiny \color{red} $\downarrow$ 13.07\%}}$    & $13.76_{\text{\tiny \color{red} $\downarrow$ 17.80\%}}$ & $30.00_{\text{\tiny \color{red} $\downarrow$ 26.67\%}}$   &  $69.55_{\text{\tiny \color{red} $\downarrow$ 11.03\%}}$  & $11.17_{\text{\tiny \color{red} $\downarrow$ 4.04\%}}$ 
\\
Qwen 2.5-coder-14B-I & $6.39_{\text{\tiny \color{red} $\downarrow$ 6.99\%}}$ & $11.29_{\text{\tiny \color{red} $\downarrow$ 8.95\%}}$ & $67.77_{\text{\tiny \color{red} $\downarrow$ 12.58\%}}$ & \cellcolor{gray!30}{$90.92_{\text{\tiny \color{red} $\downarrow$ 1.76\%}}$} & $43.32_{\text{\tiny \color{red} $\downarrow$ 13.19\%}}$ & $29.35_{\text{\tiny \color{red} $\downarrow$ 7.79\%}}$ & $45.83_{\text{\tiny \color{red} $\downarrow$ 22.44\%}}$ & $95.39_{\text{\tiny \color{blue} $\uparrow$ 4.15\%}}$ & $10.14_{\text{\tiny \color{red} $\downarrow$ 31.63\%}}$ \\
Qwen 3-32B & $6.48_{\text{\tiny \color{red} $\downarrow$ 24.39\%}}$ & $17.67_{\text{\tiny \color{red} $\downarrow$ 10.85\%}}$ & $68.85_{\text{\tiny \color{red} $\downarrow$ 16.70\%}}$ & $88.34_{\text{\tiny \color{blue} $\uparrow$ 1.97\%}}$ & $49.61_{\text{\tiny \color{red} $\downarrow$ 9.22\%}}$ & $33.93_{\text{\tiny \color{red} $\downarrow$ 11.39\%}}$ & $41.66_{\text{\tiny \color{red} $\downarrow$ 9.10\%}}$ & $97.27_{\text{\tiny \color{red} $\downarrow$ 0.50\%}}$ & $19.02_{\text{\tiny \color{red} $\downarrow$ 22.75\%}}$ \\
Deepseek-R1 & \cellcolor{gray!30}{$8.77_{\text{\tiny \color{blue} $\uparrow$ 16.08\%}}$} & \cellcolor{gray!30}{$25.18_{\text{\tiny \color{red} $\downarrow$ 5.34\%}}$} & $86.67_{\text{\tiny \color{red} $\downarrow$ 0.94\%}}$ & $75.39_{\text{\tiny \color{red} $\downarrow$ 16.68\%}}$ & $59.35_{\text{\tiny \color{red} $\downarrow$ 3.13\%}}$ & $46.03_{\text{\tiny \color{red} $\downarrow$ 12.44\%}}$ & $50.00_{\text{\tiny \color{red} $\downarrow$ 16.67\%}}$ & $94.74_{\text{\tiny \color{red} $\downarrow$ 1.35\%}}$ & $21.13_{\text{\tiny \color{red} $\downarrow$ 16.32\%}}$ \\
\midrule
GPT-4o-mini & $3.88_{\text{\tiny \color{red} $\downarrow$ 34.13\%}}$ & $17.57_{\text{\tiny \color{red} $\downarrow$ 8.11\%}}$ & $68.79_{\text{\tiny \color{red} $\downarrow$ 0.91\%}}$ & $76.47_{\text{\tiny \color{red} $\downarrow$ 1.62\%}}$ & $53.81_{\text{\tiny \color{red} $\downarrow$ 2.23\%}}$ & $31.77_{\text{\tiny \color{red} $\downarrow$ 1.00\%}}$ & $33.33_{\text{\tiny \color{red} $\downarrow$ 7.42\%}}$ & $94.85_{\text{\tiny \color{red} $\downarrow$ 1.78\%}}$ & $8.92_{\text{\tiny \color{blue} $\uparrow$ 16.91\%}}$ \\
Gemini-2.5-Pro & $6.75_{\text{\tiny \color{red} $\downarrow$ 39.79\%}}$ & $20.20_{\text{\tiny \color{red} $\downarrow$ 16.67\%}}$ & \cellcolor{gray!30}{$90.63_{\text{\tiny \color{red} $\downarrow$ 1.57\%}}$} & $90.32_{\text{\tiny \color{red} $\downarrow$ 0.58\%}}$ & \cellcolor{gray!30}{$63.15_{\text{\tiny \color{red} $\downarrow$ 5.80\%}}$} & \cellcolor{gray!30}{$50.93_{\text{\tiny \color{red} $\downarrow$ 8.51\%}}$} & $37.50_{\text{\tiny \color{red} $\downarrow$ 21.58\%}}$ & \cellcolor{gray!30}{$97.35_{\text{\tiny \color{red} $\downarrow$ 0.95\%}}$} & $26.67_{\text{\tiny \color{red} $\downarrow$ 7.78\%}}$ \\
Claude-3.7-sonnet & $7.93_{\text{\tiny \color{red} $\downarrow$ 35.74\%}}$ & $22.45_{\text{\tiny \color{red} $\downarrow$ 25.04\%}}$ & $84.37_{\text{\tiny \color{red} $\downarrow$ 2.54\%}}$ & $86.26_{\text{\tiny \color{red} $\downarrow$ 3.38\%}}$ & $58.95_{\text{\tiny \color{red} $\downarrow$ 2.50\%}}$ & $47.03_{\text{\tiny \color{red} $\downarrow$ 3.39\%}}$ & \cellcolor{gray!30}{$76.00_{\text{\tiny \color{red} $\downarrow$ 13.64\%}}$} & $91.20_{\text{\tiny \color{blue} $\uparrow$ 1.68\%}}$ & \cellcolor{gray!30}{$32.84_{\text{\tiny \color{red} $\downarrow$ 0.27\%}}$} \\
\bottomrule
\end{tabular}
}
\end{table}

%% file: Tab/RQ3.tex
\begin{table}\small
\centering
\caption{Performance on related tasks after adding metadata as one of the evidence combined with CIRs to generate behavior reports. Blue and red indicators highlight differences compared to results derived from Table~\ref{tab: rq1} All results are presented as percentages.}
\label{tab: rq3}
\resizebox{\linewidth}{!}{
\begin{tabular}{c|c|ccc|cc|ccc} 
\toprule
\multirow{2}{*}{Models} & Task1 & \multicolumn{3}{c|}{Task2}             & \multicolumn{2}{c|}{Task3}                                                     & \multicolumn{3}{c}{Task4}        \\ 
\cmidrule(lr){2-10}
                  & $\Delta S$     & AE-F1  & EAS  & SAS  & B-F1 & RQ & FPCR  & TPMR  & $F1_c$   \\ 
\midrule
Llama 3.1-8B-I   &   $0.90_{\text{\tiny \color{red} $\downarrow$ 36.17\%}}$ &   $6.27_{\text{\tiny \color{red} $\downarrow$ 9.65\%}}$    & $55.57_{\text{\tiny \color{red} $\downarrow$ 4.65\%}}$  & $44.63_{\text{\tiny \color{red} $\downarrow$ 35.43\%}} $           & $40.22_{\text{\tiny \color{blue} $\uparrow$ 10.46\%}} $     & $18.08_{\text{\tiny \color{blue} $\uparrow$ 8.00\%}} $                       & $50.00_{\text{\tiny \color{blue} $\uparrow$ 22.22\%}}  $        & $90.74_{\text{\tiny \color{blue} $\uparrow$ 16.08\%}}$ & $6.57_{\text{\tiny \color{red} $\downarrow$ 43.56\%}} $  \\
Qwen 2.5-coder-14B-I  &  $6.26_{\text{\tiny \color{red} $\downarrow$ 9.65\%}}$  &  $12.00_{\text{\tiny \color{red} $\downarrow$ 3.23\%}}$   & $78.18_{\text{\tiny \color{blue} $\uparrow$ 0.85\%}}$  & $79.42_{\text{\tiny \color{red} $\downarrow$ 14.19\%}} $           & $48.73_{\text{\tiny \color{red} $\downarrow$ 2.34\%}} $      & $30.79_{\text{\tiny \color{red} $\downarrow$ 3.27\%}} $                       & $54.17_{\text{\tiny \color{red} $\downarrow$ 8.33\%}} $          & $90.10_{\text{\tiny \color{red} $\downarrow$ 1.63\%}}$ &  $14.33_{\text{\tiny \color{red} $\downarrow$ 3.37\%}}$  \\
Qwen 3-32B    &     $6.24_{\text{\tiny \color{red} $\downarrow$ 27.19\%}}$    &  $18.11_{\text{\tiny \color{red} $\downarrow$ 8.63\%}}$   & $82.59_{\text{\tiny \color{red} $\downarrow$ 0.07\%}}$ & $83.51_{\text{\tiny \color{red} $\downarrow$ 3.60\%}}$          & $54.32_{\text{\tiny \color{red} $\downarrow$ 0.60\%}}$        & $37.34_{\text{\tiny \color{red} $\downarrow$ 2.48\%}}$       & $46.84_{\text{\tiny \color{blue} $\downarrow$ 2.20\%}}$            & $96.44_{\text{\tiny \color{red} $\downarrow$ 1.35\%}}$ & $29.01_{\text{\tiny \color{blue} $\uparrow$ 17.83\%}}$   \\
Deepseek-R1    &    \cellcolor{gray!30}{$8.43_{\text{\tiny \color{red} $\downarrow$ 19.33\%}}$}     &  $25.83_{\text{\tiny \color{red} $\downarrow$ 2.89\%}}$   & $86.09_{\text{\tiny \color{red} $\downarrow$ 1.60\%}}$ & $84.32_{\text{\tiny \color{red} $\downarrow$ 6.81\%}} $   & $61.43_{\text{\tiny \color{blue} $\uparrow$ 0.26\%}} $         & $46.23_{\text{\tiny \color{red} $\downarrow$ 12.06\%}} $           & $91.30_{\text{\tiny \color{blue} $\uparrow$ 52.17\%}} $          & $96.79_{\text{\tiny \color{blue} $\uparrow$ 0.78\%}}$ & $30.01_{\text{\tiny \color{blue} $\uparrow$ 18.85\%}}$      \\ 
\midrule
GPT-4o-mini    &   $3.32_{\text{\tiny \color{red} $\downarrow$ 43.63\%}}$      &  $15.44_{\text{\tiny \color{red} $\downarrow$ 19.25\%}}$   & $69.38_{\text{\tiny \color{red} $\downarrow$ 0.00\%}}$  & $73.82_{\text{\tiny \color{red} $\downarrow$ 5.03\%}} $         & $56.32_{\text{\tiny \color{blue} $\downarrow$ 2.33\%}} $  & $28.76_{\text{\tiny \color{red} $\downarrow$ 10.38\%}} $        & $68.00_{\text{\tiny \color{blue} $\uparrow$ 88.89\%}}  $           & $97.42_{\text{\tiny \color{blue} $\uparrow$ 0.88\%}} $& $12.54_{\text{\tiny \color{blue} $\uparrow$ 64.35\%}} $   \\
Gemini-2.5-Pro  &   $6.22_{\text{\tiny \color{red} $\downarrow$ 44.51\%}}$   &   \cellcolor{gray!30}{$26.38_{\text{\tiny \color{blue} $\uparrow$ 8.83\%}}$}  & \cellcolor{gray!30}{$87.86_{\text{\tiny \color{red} $\downarrow$ 4.58\%}}$}  &   $76.88_{\text{\tiny \color{red} $\downarrow$ 15.38\%}} $     &     \cellcolor{gray!30}{$68.57_{\text{\tiny \color{blue} $\uparrow$ 2.28\%}}$}  &   \cellcolor{gray!30}{$54.12_{\text{\tiny \color{red} $\downarrow$ 2.78\%}}$} &   $83.33_{\text{\tiny \color{blue} $\uparrow$ 74.26\%}} $  & \cellcolor{gray!30}{$99.57_{\text{\tiny \color{blue} $\uparrow$ 1.31\%}} $} & $33.72_{\text{\tiny \color{blue} $\uparrow$ 16.60\%}} $ \\
Claude-3.7-sonnet & $6.81_{\text{\tiny \color{red} $\downarrow$ 44.81\%}}$    &  $24.04_{\text{\tiny \color{red} $\downarrow$ 19.73\%}}$   & $84.27_{\text{\tiny \color{red} $\downarrow$ 2.66\%}}$  & $\cellcolor{gray!30}{86.00}_{\text{\tiny \color{red} $\downarrow$ 3.67\%}}$         & $60.78_{\text{\tiny \color{blue} $\uparrow$ 0.53\%}}$             & $48.48_{\text{\tiny \color{red} $\downarrow$ 0.41\%}}$             & \cellcolor{gray!30}{$100.00_{\text{\tiny \color{blue} $\downarrow$ 13.64\%}}$}         &  $90.52_{\text{\tiny \color{blue} $\uparrow$ 0.93\%}}$  &  \cellcolor{gray!30}{$34.83_{\text{\tiny \color{blue} $\uparrow$ 5.77\%}}$} \\
\bottomrule
\end{tabular}
}
\end{table}

%% file: Tab/category_metric.tex
\begin{table}\small
\centering
\caption{\diff{Category-level failure rates, dominant misclassification targets, B-F1, RQ, and dominant-behavior detection rates ($B_{dom}$) for Claude, DeepSeek, and Gemini across malware categories.}}
\label{tab:misclassification}
\resizebox{1.0\linewidth}{!}{
\begin{tabular}{c|ccc|c|ccc|ccc|ccc} 
\toprule
\multirow{2}{*}{Category} & Claude & Deepseek & Gemini        & \multicolumn{1}{c|}{\multirow{2}{*}{\begin{tabular}[c]{@{}c@{}}Misclassified \\Target\end{tabular}}} & Claude & Deepseek & Gemini & Claude & Deepseek & Gemini & Claude & Deepseek & Gemini \\
\cmidrule(lr){2-4} \cmidrule(lr){6-14} 
                          & \multicolumn{3}{c|}{Failure Rate} & \multicolumn{1}{c|}{}                                                                                & \multicolumn{3}{c|}{B-F1}  & \multicolumn{3}{c|}{RQ}    & \multicolumn{3}{c}{$B_{dom}$}  \\ 
\midrule
Adware                    & 52.00  & 45.00    & 42.00         &   Spyware   & 58.51  & 61.89    & 66.90  & 39.01  & 39.36    & 30.32  & 28.00 & 32.00 & 48.00 \\
Banker                    & 73.77  & 86.89    & 77.05         &   Trojan    & 57.62  & 56.14    & 49.79  & 46.44  & 43.76    & 47.20  & 21.31 & 13.11 & 24.59  \\
Ransomware                & 25.71  & 40.00    & 28.57         &   Spyware   & 64.70  & 59.60    & 69.89  & 40.71  & 43.82    & 47.24 &  71.43 & 60.00 & 74.28\\
Rootkit                   & 100.00 & 96.88    & 100.00        &  Trojan & 62.81  & 66.67    & 71.98  & 55.20  & 59.26    & 60.18  & 31.25 & 34.38 & 46.88 \\
Spyware                   & 22.64  & 20.75    & 49.06         &   Trojan   & 60.68  & 62.80    & 68.89  & 35.31  & 37.51    & 34.02  & 84.91 & 86.79 & 96.23 \\
Trojan                    & 48.15  & 59.26    & 62.72         &    Spyware  & 60.35  & 61.56    & 65.33  & 42.67  & 39.78    & 47.67  & / &/ &/  \\
\bottomrule
\end{tabular}
}
\end{table}

%% file: Tab/case_taxonomy.tex
\begin{table}
\centering
\caption{\diff{Failure mode taxonomy with diagnostic attributes and prevalence. Prevalence (\%) is estimated from 350 reports (50 malware samples $\times$ 7 models). Modes are not mutually exclusive.}}
\label{tab:failure_modes}
\resizebox{1.0\linewidth}{!}{
\begin{tabular}{cccccccc}
\toprule
ID   & Failure Mode               & Evidence & Semantic  & Chain Completeness & Attribution & Error Outcome & Prevalence(\%) \\ 
\midrule
FM-1 & Decisive Evidence Missing  &   \ding{55}       &  \ding{55}        &    \ding{55}          &      \ding{55}       &         under detection     & 34.48      \\
FM-2 & Evidence Misinterpretation &    \ding{51}      &  \ding{55}        &        \ding{55}      &   \ding{55}          &        under detection       &   3.74 \\
FM-3 & Attack-Chain Composition   &    \ding{51}      &       \ding{51}   &      \ding{55}        &    \ding{55}         &         under detection     &   23.56  \\
FM-4 & Threat Attribution         &     \ding{51}      &         \ding{51}     &      \ding{51}        &    \ding{55}        &      misattribution  & 35.63          \\
FM-5 & Unsupported Extrapolation  &     \ding{51}     &      \ding{51}    &       \ding{51}       &     \ding{51}        &        over detection    & 2.59      \\
\bottomrule
\end{tabular}
}
\end{table}

%% file: Tex/Conclusion.tex
\section{Conclusion}
\label{sec:conclusion}
\diff{This paper presents \model, a diagnostic evaluation framework for fine-grained malware behavior auditing with large language models. \model\ addresses three central challenges in auditing evaluation: the scarcity of expert-written behavior ground truth, the noise introduced by large application codebases, and the need to verify whether behavioral claims are supported by concrete code evidence. To this end, \model\ constructs context-driven intermediate representations to provide bounded and traceable code contexts, and maps both model outputs and expert reports into a shared evidence, behavior, and verdict space through constrained structural reasoning. Built on this common representation, \model\ operationalizes malware auditing as four analyst-aligned tasks, enabling stage-wise evaluation from function prioritization to sample discrimination. Applying \model\ to seven mainstream LLMs reveals persistent gaps in evidence attribution, attack-chain synthesis, and malware category understanding. These findings highlight the need for future LLM-based and agentic auditing systems to move beyond plausible summaries toward evidence-grounded and analyst-verifiable reasoning.}

%% file: Tex/DataAvailability.tex
\section{Data Availability}
The code and dataset for \model\ are publicly available at
\url{https://github.com/ZhengXR930/MalEval}.